\documentclass{article}
\title{A Steganographic Design Paradigm for General Steganographic Objectives}
\author{Aubrey Alston (ada2145@columbia.edu)}

\usepackage[left=2.5cm,top=3cm,right=2.5cm,bottom=3cm,bindingoffset=0.5cm]{geometry}

\usepackage{fancyhdr}
\date{}

\usepackage{amsmath}
\usepackage{amssymb}
\usepackage{MnSymbol}
\usepackage{graphicx}
\usepackage{float}

\usepackage{algorithm}
\usepackage{algpseudocode}
\usepackage{algorithmicx}
\usepackage{program}

\begin{document}

\maketitle

\section{Overview}

Steganography is the task of concealing a message within a medium such that the presence of the 
hidden message cannot be detected.  Beyond the standard scope of private-key steganography, steganography 
is also potentially interesting from other perspectives; for example, the prospect of steganographic parallels to 
components in public-key cryptography is particularly interesting.  In this project, 
I begin with an exploration of public-key steganography, 
and I continue by condensing existing work into a unifying design paradigm that (a) admits provably secret 
public- and private-key constructions and (b) provides for a conceptual decoupling of channel considerations 
and steganographic goals, ultimately implying both universal constructions and constructions with channel-specific optimizations.  

This work is by-and-large a survey of applications of this paradigm: 
specifically, I use the framework to achieve provably secure distributed steganography, obtain new public-key steganographic 
constructions using alternative assumptions, and give discussion of channel-specific optimizations allowed by 
cryptography as a channel and natural language channels and challenges facing practical deployment 
of steganographic systems at scale.  \footnote{This work is the report produced as a result of research performed with the Columbia university 
cryptography group.}

\tableofcontents

\section{Paper: Public-key Steganography, an Alternative to Private-key Steganography}

``Public-key Steganography'' is a work by Ahn and Hopper which explores a formal definition 
of provably secure public-key steganography \cite{BiglouPubKey}.  Unlike in the case of 
private-key steganography, public-key steganography allows for the exchange of covert 
messages without exchanging secrets.  Though an interesting prospect, the authors note that 
this goal is information-theoretically impossible; as such, 
the authors in this work attempt to provide what they claim to be the first complexity-theoretic basis for proving and 
achieving secure public-key stegosystems using standard cryptographic assumptions.
\newline\newline
To provide some motivation for their work, Ahn and Hopper begin with the standard presentation of the 
prisoners problem: Alice and Bob are attempting to covertly communicate in prison without alerting 
Ward to said covert communication, noting that the asymmetric nature of public-key steganography would not 
require an explicit secret exchange between Alice and Bob (before or after) coming to prison, unlike in the private-key setting.
\newline\newline
Towards defining steganographic security in a public-key settings, the authors provide a slightly modified presentation 
of the notion of a \textbf{channel} from \cite{BiglouPSS}.  (Initially) Rather than abstracting a \textit{channel} 
as a statistical distribution of labeled bit sequences, this work abstracts a channel as a statistical distribution 
$\mathcal{C}$ over sequences of documents $s \in D^*$.\footnote{This abstraction improves upon the original formulation given in 
\cite{BiglouPSS}, but we will see that they impose similar restrictions.}

The authors denote by $\mathcal{C}_h$ the marginal distribution on the next document conditioned on the next 
$l$ documents; similarly, they denote by $\mathcal{C}_h^l$ the marginal distribution on the next $l$ documents.
They say that a channel $\mathcal{C}$ is \textit{L-informative} if there exist constants $L > 0, b > 0, \alpha > 0$ such that 
for all $L$-length document histories $h \in D^L$, 

\[Pr_{\mathcal{C}}[h] = 0 \text{ or } H_{\infty}(C_h^{lb}) > \alpha \]

\noindent in other words, 
that the minimum entropy among all $l*b$-length document sequences in the support of $C_h^{lb}$ is some constant $\alpha$.
\newline\newline
The authors define a \textbf{public-key stegosystem} as a triple of algorithms $(KG, SE, SD)$.  $KG$ is an algorithm which 
returns as output a public-private key-pair $(PK, SK)$; $SE$ is a function $SE(PK, m, h, O)$ which takes as input a public key $PK$,
a hiddentext $m$, a message history $h$, and a channel oracle $O$ and outputs a sequence of `documents' $s = s_1,...,s_l$, the stegotext, 
from the support of $\mathcal{C}_h^l$.  $SD$ is a function $SD(SK, s, h)$ which returns the hiddentext $m$ contained within the stegotext $s$.
The correctness of a public-key stegosystem is defined by the condition

\[ \forall m \in \{0,1\}^{p(k)}, Pr_{(PK,SK)}[SD(SK,SE(PK,m,h),h) = m ] \geq 1 - \mu(k) \]

\noindent where $k$ is the security parameter and $\mu(k)$ is a negligible function.

With respect to the security of a public-key stegosystem, the authors define two settings: \textbf{chosen hiddentext secrecy} and 
\textbf{chosen stegotext secrecy}.  In the setting of \textit{chosen hiddentext secrecy}, a keypair $(PK, SK)$ is drawn according to 
security parameter $k$, and the adversary $W$ is given access to $PK$ and $\mathcal{C}_h$ for arbitrary histories $h$.  $W$ then works to 
generate a challenge history $h_W$ and message $m_W$.  $W$ is then given a document sequence $s$, where $s = SE(PK,h_W,m_W)$ or 
$s \Leftarrow C_{h_W}^{\lvert SE(PK,h_W, m_W) \rvert}$. $W$ must output a guess, 1 if he believes $s$ is the steganographically hidden 
message, 0 if he believes it was drawn according to the channel distribution $C_{h_W}$.  The advantage of an adversary is then defined as

\[ Adv(W) =  \lvert Pr_{(PK,SK)}[W^{\mathcal{C}}(PK,s=SE(\cdot,\cdot,\cdot)) = 1] - Pr_{(PK,SK)}[W^{\mathcal{C}}(PK,s \Leftarrow C_{h_W}) = 1] \rvert \]

\noindent A public-key stegosystem is considered secret under chosen hiddentext attack if, for any adversary $W$ running in time polynomial in $k$,
$Adv(W)$ is a negligible function in $k$.

\textit{Chosen stegotext secrecy} is defined to model a scenario in which an adversary $W$ is able to modify or inject stegotexts into 
the stream of communication from Alice to Bob and then uses hiddentexts output by Bob to perform steganalysis.  The setting given is essentially 
equivalent to that given for chosen hiddentext secrecy, except Alice and Bob instead maintain a public-private pair that is kept secret from Ward.
Due to the fact that the solution presented reduces to keeping steganographic public keys private within a steganographic trust architecture, and due to 
doubts about the necessity of the premise of chosen stegotext secrecy, I omit precise details [see section 1.1.2 for discussion].
\newline\newline
The specific constructions provided by Ahn and Hopper in \cite{BiglouPubKey} rely on the existence of public-key cryptography 
defined under a slightly modified definition of security.  The authors define \textit{IND-\$CPA} security as follows for a 
public-key encryption scheme: given a probabilistic scheme $H = ((PK,SK), E, D)$, define the advantage of an adversary 
$A$ as

\[Adv_H^{\$cpa}(A) = \lvert Pr_{PK}[A(PK, E_{PK}(\cdot))] - Pr_{PK}[A(PK, F(\cdot) ~ U_\{\lvert E_{PK}(\cdot) \rvert\}] \rvert\]

\noindent  (Note that $U_\{\lvert x \rvert\}$ refers to the uniform distribution over $\lvert x \rvert$-bit strings.)  Equivalently stated, 
$IND-\$CPA$ requires that ciphertexts obtained by encrypting using a public key be indistinguishable from random.  The authors 
provide and prove two \textit{IND-\$CPA} public-key encryption schemes; the first of these is a modification of RSA using 
Yung's Probabilistic Bias Removal Method (PBRM); the second is a modified manifestation of ElGamal relying on a pseudo-random 
private-key encryption scheme (i.e., a PRP).

The authors also require the existence of what they call \textit{unbiased functions} with respect to the channel distribution $\mathcal{C}$.
A function $f$ is unbiased with respect to $\mathcal{C}$ if the quantity

\[ \lvert Pr_{x\Leftarrow \mathcal{C}}[f(x) = 0] - \frac{1}{2} \rvert \]

\noindent is negligible in the security parameter of the system.

The authors then additionally rely on a basic encoding/decoding scheme which utilizes $f$ to embed uniformly distributed strings within 
documents sampled from the conditional channel distribution: 

\begin{algorithm}[H]
\caption{Basic Encoding Procedure}\label{1a}
\begin{algorithmic}[1]
\Procedure{$encode$}{$c_1,...,c_l \in \{0,1\}^l, h \in D^*, k$}
\For{i = 1...l}
\State Set $j = 0$
\While{$(j = 0 OR f(s_i) != c_i) AND (j \leq k)$}
\State $s_i \Leftarrow \mathcal{C}_h$
\State Increment $j$
\EndWhile
\State Set $h = h \mid \mid s_i$
\EndFor
\State \textbf{Return} $s_1,...,s_l$
\EndProcedure
\end{algorithmic}
\end{algorithm}

\begin{algorithm}[H]
\caption{Basic Decoding Procedure}\label{1a}
\begin{algorithmic}[1]
\Procedure{$decode$}{$s_1,...,s_l$}
\For{i = 1...l}
\State Set $c_i = f(s_i)$
\EndFor
\State \textbf{Return} $c_1,...,c_l$
\EndProcedure
\end{algorithmic}
\end{algorithm}

Finally, the authors provide a simple scheme which utilizes $f$, the above encoding/decoding procedure, and an IND\$-CPA public-key encryption 
scheme to achieve public-key steganographic encoding:

\begin{algorithm}[H]
\caption{Public-key Steganographic Encoding Procedure}\label{1a}
\begin{algorithmic}[1]
\Procedure{$encode$}{$PK, m, h$}
\State Set $c = Enc_{PK}(m)$
\State \textbf{Return} $encode(c,h,k)$
\EndProcedure
\end{algorithmic}
\end{algorithm}

\begin{algorithm}[H]
\caption{Public-key Steganographic Decoding Procedure}\label{1a}
\begin{algorithmic}[1]
\Procedure{$decode$}{$SK, s = s_1,...,s_l$}
\State Set $c = decode(s)$
\State \textbf{Return} $Dec_{SK}(c)$
\EndProcedure
\end{algorithmic}
\end{algorithm}

As a final step, the authors of \cite{BiglouPubKey} discuss the prospect of steganographic key exchange protocols.
They provide a formal definition of steganographic key exchange correctness and secrecy in the case of asynchronous single-round 
exchange protocols, and they further give a single-round protocol which is provably secure under the standard decisional 
Diffie-Hellman assumption.

The authors define a \textit{steganographic key exchange protocol} (SKEP) is a quadruple of probabilistic algorithms
$(SE_A, SE_B, SD_A, SD_B)$.  $SE_{A/B}$ should take as input a security parameter and a source of randomness and return 
as output a sequence of documents indistinguishable from $C_h^{l(k)}$ for some polynomial function $l$.  $SD_{x \in \{A,B\}}$ takes
as input a security parameter, a source of randomness $r_x$, and a sequence of documents and returns a key $K \in \{0,1\}^k$.  A SKEP
is correct if 

\[ Pr_{r_A,r_B}[ SD_A(k,r_A,SE_B(k, r_B)) = SD_B(k,r_B,SE_A(k, r_A)) ] \geq 1 - \mu(k) \]

\noindent where $\mu{k}$ is a negligible function in $k$.  A SKEP is considered secure if it is steganographically secret 
in the same setting as that for chosen hidden-text security (except where indistinguishability is now between keys as opposed to 
messages from the channel space).

\subsection{Initial Thoughts and Questions}

\subsubsection{Restrictions Placed upon Channels}

As pointed out in the report preceding this one, the formal definition of a channel given originally in 
\cite{BiglouPSS} suffers two potential flaws: (1) the reliance on fixed-size blocks and (2) strict minimum entropy 
requirements.  The formulation of a channel given in \cite{BiglouPubKey} seems to address (1) by 
abstracting the channel as being defined over a sequence of variable-length documents; however, (2) is 
still present in the requirement that all channels be $L$-informative.  With respect to (1), the issue itself may well
not be completely eliminated, as the authors in \cite{BiglouPubKey} require that channels fit requirements for some fixed-length of document sequences, 
leading again to the same concerns regarding missing opportunities to exploit structural properties in the channel to achieve higher rate.

The formal definition in \cite{BiglouPubKey} also presents new difficulty in defining security settings for steganography.
Since there is no mention or use of message length in this new formulation, there is technically no explicit bound on the 
number of bits read by an adversary (with respect to sampling the channel); this leads to difficulty in presenting precise
descriptions of adversaries from a complexity-theoretic perspective.  For further discussion of these issues, please refer to 
sections 2.1.1 and 3.2 of the previous report, where they are explored in detail and an alternate channel formulation 
(explicitly respecting complexity-theoretic definitions of adversaries) is presented.

It is also important to note that the channel formulation in \cite{BiglouPubKey} eliminates the incorporation of message/document parameters 
from \cite{BiglouPSS}, also eliminating the possibility of explicitly including (channel-specific) side-channel considerations into the construction 
of schemes.

\subsubsection{Necessity of Chosen Stegotext Security}

While the definition of \textit{chosen stegotext secrecy} is sound for the desired setting, I question the premise on the grounds of necessity.
To recap, the authors state that chosen hiddentext secrecy is defined with respect to a setting in which Ward injects new or modified stegotexts 
into the communication stream from Alice to Bob and then views hiddentexts output by Bob.  Is there any useful or realistic steganographic scenario 
in which Bob would ever output hiddentexts in response to secret messages passed by Alice?  Is there any way in which doing so would not provide anecdotal evidence of the use of steganography (or, in fact, explicitly defeat the purpose of using steganography due to taking actions easily distinguishable from 
the expected distribution $\mathcal{C}_h$)?

I would pose that a public-key stegosystem would be sufficiently secure under (a) a guarantee of chosen hiddentext secrecy, (b) a guarantee 
of chosen \textit{ciphertext} security\textit{In line with the authors' method of building stegosystems using existing public-key systems, 
this could be done by simply constructing a stegosystem using a CCA2-secure public-key system.}, and (c) the use of (either the same or another) chosen-hiddentext secret stegosystem to respond to hidden messages.  (a) should guarantee that Ward remains unaware of the use of steganography; (b) guarantees that private keys remain private among those aware of the use of steganography (thus allowing repudiation in the case that Ward can guess or influence hiddentexts); (c) keeps communication secret in continuity (even when the second party responds).  Another alternative could be 
to only use a public-key stegosystem only in an authentication step of an authenticated steganographic key exchange protocol and then use the exchanged key 
to communicate using a private-key stegosystem (which must be, of course, steganographically secret).

It is interesting to consider the sort of concern posed by the concept of chosen stegotext security: let's say that the adversary $W$ somehow learns 
Bob's public key.  $W$ can now clearly submit stegotexts to Bob or Alice without necessarily being sure that the two are using steganography, but 
neither may respond (publicly) without confirming the suspicion of $W$.  Extrapolating this scenario to a distributed system of communicating parties, 
we see immediately a need for steganographic authentication, trust management, and public-key infrastructure that guarantees secrecy as well as the 
standard guarantees of cryptographic PKIs.

\subsubsection{Public-key Steganography and the Random Oracle Model}

It seems that we can actually relax the requirement of unbiased functions in the constructions of \cite{BiglouPubKey} to obtain a result 
that random-indistinguishable public-key cryptography implies public-key steganography in the random oracle model.

Rather than expect $f$ to be an unbiased function over the support of $\mathcal{C}_h$, we may simply view $f$ as a 1-bit random oracle 
over the support of polynomial-length document sequences.  The generic encoding and decoding procedures would then be as follows:

\begin{algorithm}[H]
\caption{Basic Encoding Procedure}\label{1a}
\begin{algorithmic}[1]
\Procedure{$encode$}{$c_1,...,c_l \in \{0,1\}^l, h \in D^*, k$}
\For{i = 1...l}
\State Set $j = 0$
\While{$(j = 0 OR f(g(s_1,...,s_i)) != c_i) AND (j \leq k)$}
\State $s_i \Leftarrow \mathcal{C}_h$
\State Increment $j$
\EndWhile
\State Set $h = h \mid \mid s_i$
\EndFor
\State \textbf{Return} $s_1,...,s_l$
\EndProcedure
\end{algorithmic}
\end{algorithm}

\begin{algorithm}[H]
\caption{Basic Decoding Procedure}\label{1a}
\begin{algorithmic}[1]
\Procedure{$decode$}{$s_1,...,s_l$}
\For{i = 1...l}
\State Set $c_i = f(g(s_1,...,s_i))$
\EndFor
\State \textbf{Return} $c_1,...,c_l$
\EndProcedure
\end{algorithmic}
\end{algorithm}

\noindent (Above, $g(s_1,...,s_j)$ corresponds to some function of a sequence of messages, perhaps $g(s_1,...,s_j) = j \mid \mid s_j$ or 
$g(s_1,...,s_j) = s_1 \mid \mid ... \mid \mid s_j$.

Under this encoding scheme, in the case of a statistically uncharacterized channel, we can potentially implement arbitrary 
public-key steganographic primitives by simply implementing $f$ as any of the current candidates (e.g. the first bit of SHA-3); for well-characterized 
(and well-structured) channels, we may rely instead on some function $f$ which is a genuinely unbiased function with respect tot he channel 
distribution.

\subsubsection{Definition of Security for Steganographic Key Exchanges}

The authors of \cite{BiglouPubKey} note that their definition of a steganographic key exchange (and definition of security) is limited to the 
asynchronous, single-round case.  It seems worthwhile to pursue a definition which suffices for multi-round protocols.  (I attempt to provide such a 
definition and security setting in a later section.)

\section{Operate-Embed-Extract: A Stegosystem Design Paradigm}

\cite{BiglouPubKey} and \cite{BiglouPSS} respectively show the existence of private-key and public-key steganography in arbitrary channels 
meeting minimum entropy requirements.  Both works provide multiple provably secret constructions for steganographic equivalents of 
private-key and public-key encryption based upon different assumptions, but, interestingly, each of these constructions (1) seems to 
intuitively fit into a common, unstated pattern while (2) also somehow being excessively restrictive in their pursuit of `universal' steganography 
to the point of preventing use of channel-specific qualities for the sake of either efficiency or security.

In this section, I attempt to consolidate the common methods of the steganographic constructions of \cite{BiglouPubKey} and \cite{BiglouPSS} 
into a single abstract paradigm for stegosystem design.  This paradigm, which I call \textit{operate-embed-extract}, serves to provide 
a systematic framework for designing and proving the security of stegosystems both universally and for specific channels.

\subsection{Paradigm Description}

The \textit{operate-embed-extract} paradigm is applicable to the design of constructions serving the purpose of a steganographic objective.  
The steganographic objectives applicable to the paradigm are those which may be stated as an equivalent cryptographic objective having 
the additioinal constraint that any output produced must be indistinguishable from some channel distribution $\mathcal{C}$.
\newline\newline
A steganographic construction $S$ in the \textit{operate-embed-extract} paradigm is composed of at least two probabilistic algorithms
$(S, S^{-1})$ which make use of three components: a set of external operations $F_{ext} = \{ OPERATE'_1(<r_{public}, r_{private}>,o),...\}$, a 
set of internal operations $F_{int} = \{ OPERATE_1^{(int)}(<r_{public}, r_{private}>,o),...\}$, and two other functions, 
$EMBED(\mathcal{C},h,t~\mathcal{D}), EXTRACT(\mathcal{C},h,c_1,...,c_{l(k)})$.

$OPERATE(\cdot,\cdot) \in F_{ext}$ takes as input a source of public randomness, a source of private randomness, and an objective string $o$; 
$OPERATE$ uses $<r_{public}, r_{private}>$ to apply some transformation to $o$ and returns output 
$d$ computationally indistinguishable from distribution $\mathcal{D}$.  All public $OPERATE$ functionalities must also guarantee 
the desired cryptographic properties of the construction.

$OPERATE(\cdot,\cdot) \in F_{int}$ is simply any function of a source of randomness and an objective string $o$.  These functions need not 
perform any specific purpose and are instead defined as needed by the stegosystem.

$EMBED$ takes as input a channel $\mathcal{C}$, a history $h$, and an element $t$ drawn from an \textit{input distribution} $\mathcal{D}$ and returns a sequence
$c_1,...,c_{l(k)}$ from the support of $\mathcal{C}_h^{l(k)}$.  $EXTRACT$ takes as input a channel $\mathcal{C}$, a history $h$, and 
a sequence of covertexts $c_1,...,c_{l(k)}$ and returns a message from some message space.
\newline\newline
A steganographic construction $S=(T,T^-1)$ in this paradigm using would then be structured as follows: 

\begin{algorithm}[H]
\begin{algorithmic}[1]
\Procedure{$T$}{$h,r_{public}, r_{private}, m \in \mathcal{M}$}
\State Some sequence of interleaved external and internal operations $\in F_{ext}$ and $\in F_{int}$, obtaining $m'$.
\State $m'' = F(r_{public}, r_private, m')$ for some $F \in F_{ext}$ returning $m''$ from the support of $\mathcal{D}$.
\State \textbf{return} $EMBED(\mathcal{C},h,m'')$
\EndProcedure
\end{algorithmic}
\end{algorithm}

\begin{algorithm}[H]
\begin{algorithmic}[1]
\Procedure{$T^-1$}{$h,r_{public}, r_{private}, <c_1,...,c_{l(k)}>$}
\State $m'' = EXTRACT(\mathcal{C},h,<c_1,...,c_{l(k)}>)$
\State Some sequence of interleaved internal operations $\in F_{int}$, obtaining $m$ from $m''$.
\State \textbf{return} $m$
\EndProcedure
\end{algorithmic}
\end{algorithm}

\subsection{A Proof Framework for Steganographic Objectives}

The \textit{operate-embed-extract} paradigm also gives a simplified and streamlined proof framework for steganographic 
constructions, both those defined with respect to specific channels and with respect to those defined universally.
\newline\newline
I claim that, for any steganographic objective defined under the given paradigm, only the following need be proved to prove 
security:

\begin{enumerate}
\item{The output of $EMBED(\mathcal{C}, h, x)$ is indistinguishable from $\mathcal{C}_h^{l(k)}$ when $x$ is drawn according to $\mathcal{D}$.}
\item{The final output of the last-invoked $F \in F_{ext}$ satisfies all necessary cryptographic objectives.}
\item{The final output of the last-invoked $F \in F_{ext}$ is indistinguishable from $\mathcal{D}$.} 
\end{enumerate}

\noindent (Note: the scheme must also be shown to be correct in order to be valid.)
\newline\newline
Justification of this claim is simple: say that we prove that $F$ satisfies all cryptographic requirements.  Now say that 
we prove that the output of $F$ is indistinguishable from $\mathcal{D}$.  If we are using an $EMBED(\cdot,\cdot,\cdot)$ which is 
indistinguishable from $\mathcal{C}_h^{l(k)}$ given $x \sim \mathcal{D}$, then we have that $EMBED(\cdot,\cdot,F(\cdot,\cdot))$ is 
also indistinguishable from $\mathcal{C}_h^{l(k)}$.  Since the scheme (a) satisfies the necessary cryptographic requirements and 
(b) is indistinguishable from the channel distribution, the scheme is secure under the given steganographic secrecy setting.

\subsubsection{Implications}

There are two primary theoretical implications of this paradigm: (1) separation of the design of cryptographic functionality from 
channel embedding and (2) the ability to adapt stegosystems proved to be secure in a universal model in a channel-specific manner 
without affecting security.

With respect to (1), we see that this separation comes from the fact that $F_{int}$ and $F_{ext}$ are not defined with respect to a channel,
whereas all channel-specific operations are exclusive to $EMBED/EXTRACT$.  Given a steganographic objective, then, we may design the 
procedures for $F_{ext}$ and $F_{int}$ without any concern given to the channel in which the stegosystem will be applied (instead shelving 
that concern for the design of $EMBED/EXTRACT$).

Moreover, this separation leads us to (2): by defining a universal $EMBED/EXTRACT$ procedure for a class of channel (or perhaps all of them),
design of a sufficient $F_{int}$ and $F_{ext}$ procedure for any objective immediately yields a steganographically secure construction in all of 
those channels. Further, we retain the freedom to later modify $EMBED/EXTRACT$ (perhaps to achieve greater efficiency in the channel) 
at a finer granularity at a later time without affecting security (so long as the modification obeys the requirements of $EMBED/EXTRACT$, of course.)
\newline\newline
There also exists a significant benefit from a systems/implementation standpoint: the modularity of this framework increases the ease 
involved in implementing a diverse array of stegosystems, potentially in shared channels.  Fix a set of channels over which we plan to operate
and a distribution $\mathcal{D}$.  We now need only to implement $EMBED/EXTRACT$ for these channels once, and then we may implement 
any number of stegosystems for any number of steganographic objectives using these channels by simply implementing the proper $F_{priv}$ and 
$F_{pub}$ procedures which utilize $EMBED/EXTRACT$.  Use of such a framework immediately bolsters the practical utility of STEG-MQ (allowing for 
things such as shared channel steganography or even simply simple selection between different objectives).

\subsection{A Pseudo-Universal EMBED/EXTRACT Procedure in the Random Oracle Model}

As previously discussed, in the \textit{operate-embed-extract} framework,
design of a satisfactory $F_{int},F_{ext}$ procedure immediately yields secure steganographic constructions for our objective 
in as many channels as are covered by $EMBED/EXTRACT$.  In this section, I re-frame an existing method described in \cite{BiglouPubKey} 
and \cite{BiglouPSS} to obtain an $EMBED/EXTRACT$ procedure which applies to any channel having entropy bounded from below by a constant 
(or, equivalently, maximum probability over any element of the support $\mathcal{C}_h$ bounded from above by a constant).

Fix $\mathcal{D}$ to be the uniform distribution over binary strings of length $p(k)$ (where $k$ is the security parameter).  Assume the existence
of a function $f(\mathcal{C},h,x)$ which, given any $\mathcal{C}$ and $h$, acts as a random oracle mapping $x$ to $\{0,1\}$ without bias 
($Pr[f(\mathcal{C},h,x) = 0] = \frac{1}{2}$).  Assume also that $EMBED$ has access to an oracle which allows it to sample from 
$\mathcal{C}_h$ for arbitrary histories $h$.  Our EMBED/EXTRACT procedure is as follows:

\begin{algorithm}[H]
\begin{algorithmic}[1]
\Procedure{$EMBED$}{$\mathcal{C}, h, x ~ U(\{0,1\}^{p(k)})$}
\For{$i = 1,...,\lvert x \rvert}$
\State $t = 0$
\While{$t < k$}
\State $c_i \Leftarrow \mathcal{C}_h^{1}$
\If $f(\mathcal{C},h,c_i) = x_i$
\State \textbf{break}
\EndIf
\EndWhile
\State $h = h \mid \mid c_i$
\EndFor
\State \textbf{return} $c1,...$
\EndProcedure
\end{algorithmic}
\end{algorithm}

\begin{algorithm}[H]
\begin{algorithmic}[1]
\Procedure{$EXTRACT$}{$\mathcal{C},h,<c_1,...,c_m>$}
\State \textbf{return} $m = f(\mathcal{C},h,c_1) \mid \mid ... \mid \mid f(\mathcal{C},h\mid\mid ... \mid\mid c_{m-1}, c_m)$
\EndProcedure
\end{algorithmic}
\end{algorithm}

\subsubsection{Proof of Indistinguishability}

The proof that the output of $EMBED$ is indistinguishable from the channel distribution is direct.  Let $C_1,...,C_m$ be random variables
whose realizations are messages from the support of $\mathcal{C}_h^{\Rightarrow(m)}$  Let $b_1,...,b_m$ be the bits of the input $x$ to embed.  
Consider any individual $C_i$.  The probability that $C_i = c$ is precisely the probability that $c$ is drawn in a trial where $f(c)=b_i$:

\begin{align*}
Pr[C_i = c \mid b_i ] &= \frac{Pr_{\mathcal{C}_h}[c] Pr[f(c) = b_i]}{Pr[b_i]} \\
&= \frac{Pr_{\mathcal{C}_h}[c] \frac{1}{2}}{Pr[b_i]} & \text{(f acts as random oracle)} 
\end{align*}

We may similarly extend this statement to the complete output: 

\begin{align*}
Pr[C_1 = c_1 ... C_m = c_m \mid b_1 ... b_m] &= \frac{Pr_{\mathcal{C}_h^{\Rightarrow(m)}}[c_1,...,c_m] Pr[f(c_1) = b_1,...,f(c_m)=b_m]}{Pr[b_1...b_m]} \\
&= \frac{Pr_{\mathcal{C}_h^{\Rightarrow(m)}}[c_1,...,c_m] \frac{1}{2}^{m}}{Pr[b_1...b_m]} & \text{(f acts as random oracle)} 
\end{align*}

We thus see that $Pr[C_1 = c_1 ... C_m = c_m \mid b_1 ... b_m]$ is $Pr_{\mathcal{C}_h^{\Rightarrow(m)}}[C_1 = c_1 ... C_m = c_m]$ when 
$x$ is drawn uniformly at random.

\subsubsection{Note on Correctness}

We note that, with negligible probability, the given EMBED/EXTRACT may be incorrect in one or more bits.  This negligibility, however, applies
only to channels meeting a minimum entropy assumption: if at any point the entropy of $\mathcal{C}_{h\mid\mid...c_i}$ is not constant and $ > 0$, 
the probability of failure is no longer necessarily bounded by a negligible function.

\subsubsection{An Entropy-Adaptive Universal EMBED/EXTRACT Procedure for Uncharacterized Channels}

We can further improve the utility of the previously given EMBED/EXTRACT procedure if we assume that $EXTRACT$ also has access to a channel oracle.
This improvement is based on the following observation: if the entropy in the channel is too low for $EMBED$, then there will not exist a $c_i$ that 
may feasibly be sampled such that $f(c_i) = b_i$.
\newline\newline
In order to capitalize on this advantage, $EMBED$ would function exactly as before, except before accepting a message from the support of the 
channel, it would verify that, within $k$ iterations, it is able to obtain $c_k$ $c_j$ such that one is mapped by $f$ to 1 and one is mapped by 
$f$ to 0.  If this is possible, then it performs as usual.  If this is not possible, then $EMBED$ just samples and uses any $c$ from the support 
of the channel, with the expectation that $EXTRACT$ will be able to come to the conclusion that it should skip it.

Likewise, $EXTRACT$ would function exactly as before, except it would perform a similar check before using the bit $f(c_i)$: it would verify that 
it is able to obtain $c_k$, $c_j$ such that one is mapped by $f$ to 1 and one is mapped by $f$ to 0 by the oracle.  If this is not possible, it skips 
the current $c_i$.  If it is possible, then it uses the current $c_i$.  

Note that the error probability for this scheme is now negligible, even for channels with entropy not bounded by a constant: if the channel 
does have high entropy, then the probability that $EMBED$ is not able to obtain covertexts mapped to both 1 and 0 is negligible; if the channel 
has low entropy, then the probability that $EMBED$ is able to obtain such a pair is negligible.  Similarly, if the channel has high entropy, 
$EXTRACT$ will err and skip a covertext only with negligible probability; if the channel has low entropy, then the probability that $EXTRACT$ 
attempts to decode using the ciphertext is negligible.
\newline\newline
We note that, in a situation in which use of this EMBED/EXTRACT mechanism is necessary, it's nearly certain that any stegosystem utilizing it 
will achieve only a low rate rate (very few covertexts may be usable) of hidden information.  The existence of this strategy, however, 
does enable a few fringe use cases that may potentially prove useful, and it extends the scope of our `universal' approach to steganography.

\section{Multi-channel (Distributed) Steganography: Achieving a New Steganographic Security Objective using Operate-Embed-Extract}

Work exists \cite{DistSteg} which attempts to `increase' secrecy by effectively splitting messages between multiple mediums.  When this method 
is suggested, the intuitive motivation seems to be that splitting messages between mediums somehow decreases the chance of detection.  In the specific 
case of \cite{DistSteg}, the authors attempt to work towards a formal model which ultimately (a) relies on methods known to be vulnerable (specifically
LSB steganography in images) and (b) does not generalize to other channels.  Ignoring the oversights of previous work, however, distributed steganography 
seems to be an interesting task in that it may allow us to develop systems which use the knowledge/accessibility of elements of a general `environment' 
to provide security guarantees.

In this section, I broadly explore a formal basis for such a distributed form of steganography.  Specifically, I 
(1) provide a formal definition of a distributed stegosystem, (2) provide two equivalent security settings for the 
provable security of a distributed stegosystem, and (3) provide a provably secret constructions for these settings.
Furthermore, I demonstrate the utility of the \textit{operate-embed-extract} paradigm by using it to design these constructions.

\subsection{Notation}

This section makes use of non-standard notation for the sake of ease of exposition.  Let $A$ be a set of ordered element-tuple pairs of the form 
$A = \{ (x_1, <e_1^{(1)},...>), (x_m, <e_1^{(m)},...>)\}$.  Let $B$ be a set of elements $B=\{x_{i_1}, ..., x_{i_k}\}$.  We define $A<B>$ as 

\[ A<B> = \{(x_a,<e_1^{(a)},...>) \mid (x_a,<e_1^{(a)},...>) \in A, x_a \in B \} \]

\subsection{A Formal Definition of Distributed Stegosystems}

We define the \textbf{environment} of a distributed stegosystem be a set $[\mathcal{C}] = \{ \mathcal{C}_(1), ..., \mathcal{C}_(w) \}$ of channels 
in which it operates.
\newline\newline
A \textbf{distributed stegosystem} is a pair of probabilistic polynomial-time algorithms $(DSE,DSD)$.  

$DSE(r,[\mathcal{C}]_t, [h]_t, m \in \{0,1\}^l(k), e)$ takes 
as input a source of randomness $r$, a target environment $[\mathcal{C}]_t \subseteq [\mathcal{C}] = \{ \mathcal{T}_1,..., \mathcal{T}_t\}$, a 
set of histories $[h]_t = \{h_1,...h_t\}$ (where $h_i$ is the history of channel $\mathcal{T}_i \in [\mathcal{C}]_t)$), a message to hide $m$ 
(having length polynomial in $k$, the security parameter), and a threshold $e \leq t$ specifying the minimum number of channels that must be accessible to 
recover $m$;  $DSE(\cdot, \cdot, \cdot, \cdot, \cdot)$ returns a set of covertext sequences, one for each channel in the target environment,
$\{(\mathcal{T}_1, <c_1^{(1)},...>),...,(\mathcal{T}_t, <c_1^{(t)},...>)\}$.

$DSD([C]_s, [h]_s, \{(\mathcal{S}_1, <c_1^{(1)},...>),...,(\mathcal{S}_s, <c_1^{(s)},...>)\})$ takes as input a visible environment 
$[\mathcal{C}]_s \subseteq [\mathcal{C}] = \{ \mathcal{S}_1,..., \mathcal{S}_s\}$, a set of histories $[h]_s = \{h_1,...,h_s\}$ (where $h_i$ 
is the history of channel $\mathcal{S}_i$), and a set of covertext sequences, one for each channel in the seen environment; $DSD(\cdot, \cdot, \cdot)$ 
returns a message $\{0,1\}^{l(k)}$.

\subsubsection{Correctness of Distributed Stegosystems}

A distributed stegosystem is considered correct if, for all input configurations satisfying $e \geq 1$, $\lvert [\mathcal{C}]_s \rvert \geq e$, 
$[\mathcal{C}]_s \subseteq [\mathcal{C}]_t$, $[h]_s \subseteq [h]_t$,

\[ Pr_r[DSD([\mathcal{C}]_s, [h]_s, DSE(r, [\mathcal{C}]_t, [h]_t, m, e)<\mathcal{C}]_s>) = m] \geq 1 - \mu(k) \]

\noindent for some negligible function $\mu(k)$.

\subsubsection{Security Setting: Open-Access Distributional Secrecy against Chosen Hiddentext Attacks}

We provide a notion of security for distributed stegosystems called 
\textit{open-access distributional secrecy against chosen hiddentext attacks} (OA-DS-CHA).  This setting is chosen to model attempted detection by 
an adversary in a chosen environment where not all communication channels visible to Alice and Bob are visible to the Warden at any given 
point in time.  In this setting, channels 
are considered \textit{open-access}, meaning visibility of cover messages is the only requirement to recover steganographically hidden information.
Security in this setting is desirable for participants Alice and Bob who believe that there exists a threshold 
number of channels visible to the Warden at the time of attempted detection but who for some reason cannot establish a pre-shared secret.

We define this setting as the following game played by a 
warden $W$ in an environment $[\mathcal{C}] = \{ \mathcal{C}_{(1)}, ..., \mathcal{C}_{(w)} \}$:

\begin{enumerate}
\item{$W$ is given oracles $O_1,...,O_w$ capable of sampling from channels $\mathcal{C}_{1},...,\mathcal{C}_{w}$ for arbitrary histories 
$h$.}
\item{$W$ is given query access to $DSE(\cdot,\cdot,\cdot,\cdot,\cdot)$.}
\item{$W$ generates a challenge set consisting of a target environment $[\mathcal{C}]_t$ ($t \geq 1$), a corresponding set of histories 
$[h]_t$, a message $\in \{0,1\}^{l(k)}$, and a threshold $e \geq 1$.}
\item{$W$ chooses a visible subset of the environment $[\mathcal{C}]_s$ of size $e-1$.}
\item{$W$ is then given one of }
\begin{enumerate}
\item{$Q_1$: $DSE(r,[\mathcal{C}]_t, [h]_t, m, e)<[\mathcal{C}]_s>$ for privately chosen (and unknown) randomness $r$, or}
\item{$Q_0$: $\{ (\mathcal{S}_1, <...> \sim \mathcal{S}_{h_{s_1}}^{p(k)}), ..., (\mathcal{S}_s, <...> \sim \mathcal{S}_{h_{s_s}}^{p(k)})\}$}
\end{enumerate}
\item{$W$ is given query access to $DSE(\cdot,\cdot,\cdot,\cdot,\cdot)$ once more.}
\item{$W$ outputs a guess: $1$ if he believes he was given $Q_1$, else $0$.}
\end{enumerate}

\noindent We define the advantage of a warden $W$ as the quantity

\[ Adv(W) = \lvert Pr_r[W(Q_1) \Rightarrow 1] - Pr_r[W(Q_0) \Rightarrow 0] \rvert \]

\noindent A distributed stegosystem $S = (DSE, DSD)$ is $DS-CHA$ secure in environment $[\mathcal{C}]$ if the advantage 
of any probabilistic polynomial-time warden is negligible in the security parameter of the system.  

By nature of the definition of DS-CHA security, a system is DS-CHA secure if and only if covertexts produced by the system 
are computationally indistinguishable from benign covertexts in visible subsets of the environment not meeting the specified size threshold $e$.

\subsubsection{Security Setting: Key-Restricted Distributional Secrecy against Chosen Hiddentext Attacks}

We also provide a stronger notion of security for distributed stegosystems called \textit{key-restricted distributional secrecy against chosen hiddentext
attacks} (KR-DS-CHA).  This setting is chosen to model detection by an adversary in a chosen environment where the adversary may see all channels 
but may recover steganographically hidden information in some (but not all) channels.  Channels in this model are \textit{key-restricted}, meaning that both visibility and key knowledge are required to recover steganographically hidden information from any given channel.  Security in this setting is desirable for participants 
Alice and Bob who believe that there exists a threshold number of channel keys which may be leaked to the adversary.

KR-DS-CHA security applies to families of distributed stegosystems indexed by a key vector, $(DSE_{\bar{K}}, DSD_{\bar{K}})$, $\bar{K} \in \{0,1\}^{kw}$.
We define this setting as the following game played by a warden $W$ in an environment 
$[\mathcal{C}] = \{ \mathcal{C}_{(1)}, ..., \mathcal{C}_{(w)} \}$ with associated channel keys 
$[\mathcal{K}] = \{ \mathcal{K}_{(C_1)}, ..., \mathcal{K}_{C_w}) \}$.

\begin{enumerate}
\item{$W$ is given oracles $O_1,...,O_w$ capable of sampling from channels $\mathcal{C}_{1},...,\mathcal{C}_{w}$ for arbitrary histories 
$h$.}
\item{$W$ is given query access to $DSE_{\bar{K}}(\cdot,\cdot,\cdot,\cdot,\cdot)$.}
\item{$W$ generates a challenge set consisting of a target environment $[\mathcal{C}]_t$ ($t \geq 1$), a corresponding set of histories 
$[h]_t$, a message $\in \{0,1\}^{l(k)}$, and a threshold $e \geq 1$.}
\item{$W$ is then given one of }
\begin{enumerate}
\item{$Q_1$: $DSE(r,[\mathcal{C}]_t, [h]_t, m, e)$ for privately chosen (and unknown) randomness $r$, or}
\item{$Q_0$: $\{ (\mathcal{C}_1, <...> \sim \mathcal{C}_{1,h_{1}}^{p(k)}), ..., (\mathcal{C}_w, <...> \sim \mathcal{C}_{w,h_{w}}^{p(k)})\}$}
\end{enumerate}
\item{$W$ chooses a subset of channels $[\mathcal{C}]_s$ of size $e-1$.}
\item{$W$ receives the $e-1$ keys corresponding to the channels in $[\mathcal{C}]_s$.}
\item{$W$ is given query access to $DSE_{\bar{K}}(\cdot,\cdot,\cdot,\cdot,\cdot)$ once more.}
\item{$W$ outputs a guess: $1$ if he believes he was given $Q_1$, else $0$.}
\end{enumerate}

\subsection{An OA-DS-CHA-secret Distributed Stegosystem}

In this section, we provide an OA-DS-CHA-secret distributed stegosystem using the operate-embed-extract stegosystem design framework. 
This scheme uses a modified variant of Shamir Secret Sharing, and the key mechanism used is interpolation.

\noindent The public parameters of this system are as follows:
\begin{enumerate}
\item{A field $\mathbb{F}_q$ for $q \geq 2^k$ (where $k$ is the security parameter), say $GF(2^k)$.  (We could also 
use a prime field, but we would then need to also augment this scheme with something like Young and Yung's probabilistic 
bias removal method \cite{YoungPBRM}.)}
\end{enumerate}

\noindent This scheme also makes use of the following set of internal functions \newline
$F_{int} = \{ P, SecretGen(r), PolyGen(r,e), PointGen(r,t), Interpolate(L, e) \}$.
$P$ is a pseudorandom permutation, and the rest are defined as follows:

\begin{algorithm}[H]
\caption{SecretGen Procedure}\label{1a}
\begin{algorithmic}[1]
\Procedure{$SecretGen$}{$r$}
\State \textbf{return} random secret $T \in \{0,1\}^k$.
\EndProcedure
\end{algorithmic}
\end{algorithm}

\begin{algorithm}[H]
\caption{PolyGen}\label{1a}
\begin{algorithmic}[1]
\Procedure{$PolyGen$}{$r, e$}
\State Choose vector $\vec{a_i}$ composed of $e-1$ values chosen uniformly at random from $\{0,1\}^k$.
\State Choose another vector $\vec{b_i}$ composed of $e-1$ values chosen uniformly at random from $\{0,1\}^k$.
\State \textbf{return} $\vec{a_i}, \vec{b_i}$
\EndProcedure
\end{algorithmic}
\end{algorithm}

\begin{algorithm}[H]
\caption{PointGen}\label{1a}
\begin{algorithmic}[1]
\Procedure{$PointGen$}{$r, t$}
\State Choose random sequence $\vec{x}$ of $t$ values uniformly at random from $\{0,1\}^k$ without replacement.
\State \textbf{return} $\vec{x}$
\EndProcedure
\end{algorithmic}
\end{algorithm}

\begin{algorithm}[H]
\caption{Interpolate (Lagrange interpolation to recover constant) }\label{1a}
\begin{algorithmic}[1]
\Procedure{$Interpolate$}{$L=\{(x_1, f(x_1)) \mid \forall i = 1 ... e \}, e$}
\State \textbf{return} $\sum_{i=1}^{e} f(x_i) \prod_{j=1\neq i}^{e} \frac{x_j}{x_j - x_i}$
\EndProcedure
\end{algorithmic}
\end{algorithm}

\noindent Our scheme also makes use of a single external function, $F_{ext} = \{ EvaluatePoint(\vec{a_i}, x) \}$.

\begin{algorithm}[H]
\caption{EvaluatePoint}\label{1a}
\begin{algorithmic}[1]
\Procedure{$EvaluatePoint$}{$p0, \vec{a_i}, x$}
\State \textbf{return} $x \mid \mid (p0 + \sum_{i=1}^{\lvert \vec{a_i} \rvert} x^i a_i)$
\EndProcedure
\end{algorithmic}
\end{algorithm}

\noindent Our scheme is thus the following (we note that all uses of randomness source $r$ are fresh; i.e., after one function 
uses random bits drawn from $r$, $r$ supplies the next with fresh bits):

\begin{algorithm}[H]
\caption{Distributed Steganographic Encoding Procedure}\label{1a}
\begin{algorithmic}[1]
\Procedure{$DSE$}{$r, [\mathcal{C}]_t, [h]_t, m \in \{0,1\}^k, 1 \leq e \leq t$}
\State Set $K = SecretGen(r)$.
\State Set $\vec{a_i}, \vec{b_i} = PolyGen(r,e)$.
\State Set $X = <x_i> = PointGen(r,t)$.
\State Set $\tilde{m} = P_K(m)$.
\For{$i$ from 1 to $t$}
\State Take $x_i \mid \mid y_i \mid \mid \tilde{y_i}$ as $EvaluatePoint(K, \vec{a_i}, x_i) \mid \mid EvaluatePoint(\tilde{m}, \vec{b_i}, x_i)_y$
\State Set $c^{(i)} = EMBED(\mathcal{C}_i, h_i, x_i \mid \mid y_i \mid \mid z_i \mid \mid \tilde{y_i})$.
\EndFor
\State \textbf{return} $\{(\mathcal{C}_i, <c^{(i)}>) \mid \forall =1...t \}$
\EndProcedure
\end{algorithmic}
\end{algorithm}

\begin{algorithm}[H]
\caption{Distributed Steganographic Decoding Procedure}\label{1a}
\begin{algorithmic}[1]
\Procedure{$DSD$}{$[\mathcal{C}]_s, [h]_s, \{ (\mathcal{S}_1, <c^{(1)}>), ..., (\mathcal{S}_s, <c^{(s)}>) \}$}
\For{$i$ from 1 to $s$}
\State Take $x_i \mid \mid y_i \mid \mid \tilde{y_i}$ as $EXTRACT(\mathcal{S}_i, h_i, c^{(i)})$.
\State Take as a point $(x_i, y_i = f(x_i))$ and append it to $L_1$.
\State Take as a point $(x_i, \tilde{y_i} = g(x_i))$ and append it to $L_2$.
\EndFor
\State Set $K = Interpolate(L_1,e)$.
\State Set $\tilde{m} = Interpolate(L_2, e)$.
\State \textbf{return} $P_K^{-1}(\tilde{m})$.
\EndProcedure
\end{algorithmic}
\end{algorithm}

Correctness follows directly from the correctness of Lagrange interpolation (for the first coefficient of a polynomial).

\subsubsection{Proof of Secrecy}

\noindent \textit{Motivation for the Proof Structure }  We give a proof of secrecy under the 
\textit{operate-embed extract} paradigm in order to show the utility of the proof 
framework it admits.  To review, in section 3.1.3, we claimed the following is sufficient to prove the security of an 
operate-embed-extract-amenable steganographic objective:

\begin{enumerate}
\item{The output of $EMBED(\mathcal{C}, h, x)$ is indistinguishable from $\mathcal{C}_h^{l(k)}$ when $x$ is drawn according to $\mathcal{D}$.}
\item{The final output of the last-invoked $F \in F_{ext}$ satisfies all necessary cryptographic objectives.}
\item{The final output of the last-invoked $F \in F_{ext}$ is indistinguishable from $\mathcal{D}$.} 
\end{enumerate}

To illustrate how we will proceed in our proof, consider an open-access distributed steganographical scheme for which there exists 
a probabilistic polynomial-time warden achieving non-negligible advantage in the OA-DS-CHA security game.  Then it is necessarily the 
case that, when playing against this scheme, the warden is able to distinguish between steganographically hidden messages and messages 
drawn from the benign channel distribution.  

Consider now the more specific case of a scheme designed using \textit{operate-embed-extract}.  Then it is necessarily the case 
that the compiled output of $EMBED(\mathcal{C}, h, x)$ in the $e-1$ selected channels is distinguishable from $\mathcal{C}_h^{l(k)}$.  It must then 
be the case that either (a) required property 1 does not hold or that (b) required property 1 does hold but that the components hidden 
are not drawn according to distribution $\mathcal{D}$.  Therefore, in the case of proving $OA-DS-CHA$ secrecy, it is sufficient to prove 
that both property 1 holds for all channels and that property 3 holds for the compiled output of 
$EvaluatePoint(K, \vec{a_i}, x_i) \mid \mid EvaluatePoint(\tilde{m}, \vec{b_i}, z_i)$ in the $e-1$ channels visible to the adversary.  
In the specific case of OA-DS-CHA secrecy, only properties 1 and 3 need to be proved.
\newline\newline
\noindent \textit{Proof }  The given scheme satisfies property 1: it assumes the use of canonical $EMBED/EXTRACT$ procedures for 
all concerned channels where $\mathcal{D}$ is fixed to the uniform distribution over binary strings of length $k$.  
As for property 3, view the inputs to the final external function of $DSE$ in the individual channels of $[\mathcal{C}]_s$ in $DSE$ (arbitrarily ordering them as $1...s$) 
as the following matrix $S$:

\[
\begin{bmatrix}
    x_1 & y_1 & \hat{y}_1 \\
    \hdotsfor{3} \\
    x_s & y_s & \hat{y}_s 
\end{bmatrix}
\]

The columns of $S$ may be interpreted as follows: column 1, $X=(x_1,...,x_s)$, is a sequence of x-coordinates of points hiding 
$K$; column 2, $Y=(y_1,...,y_s)$, is the sequence of y-coordinates of points hiding $K$ (($x_i,y_i$) is a complete share of $K$); column 3, $\hat{Y} = (\hat{y}_1,...,\hat{y}_s)$, is the sequence of x-coordinates of points hiding $P_K(m)$
($\hat{x}_i,\hat{y}_i$) is a complete share of $P_K(m)$).

As stated, property 3 is satisfied if and only if the distribution over $X \mid \mid Y \mid \mid \hat{Y}$ is indistinguishable 
from the uniform random distribution over binary strings of length $3sk$.  Consider first only the distribution of $X=x_1,...,x_s$.  
\newline\newline
\noindent \textbf{Claim 1} The distribution of $X$ and the uniform distribution $X_u$ over binary strings of length $sk$ are statistically 
indistinguishable (with respect to the security parameter $k$). 
\newline\newline
\noindent For any element in the support of $X_u$, partition it into a sequence of $s$ substrings of length $k$ as we 
do with $X$.  Since each $x_s$ is some subset of $x'_{1},...,x'_{t}$ chosen uniformly at random without replacement, we have 

\[
Pr[X=x_1,..., x_s] = \begin{cases} 
      0 & \exists i \neq j, x_i = x_j \\
      a, \frac{1}{2^{ks}} < a \leq \frac{1}{(2^k - t)^s} & \text{ else}
   \end{cases}
\]

\noindent Assuming $t$ (the number of channels in which information will be hidden) is polynomial in $k$, we have

\[
Pr[X=x_1,...,x_s] = \begin{cases} 
      0 & \exists i \neq j, x_i = x_j \\
      a, \frac{1}{2^{sk}} < a < \frac{1}{2^{s(k-1)}} & \text{ else}
   \end{cases}
\]

\noindent Consider now the statistical distance between $X$ and $X_u$ the uniform distribution over binary strings of 
length $ks$.

\begin{align*}
\Delta(X,X_u) &= \sup_{x_1,...,x_s \in \{0,1\}^{sk}}(\lvert Pr[X=x_1,...,x_s] - Pr[X_u=x_1,...,x_s] \rvert) \\
&= max(\frac{1}{2^{ks}}, \frac{1}{2^{s(k-1)}} - \frac{1}{2^{ks}}) \\
&\leq \frac{1}{2^{s(k-1)}} \\
&= \mu(k)
\end{align*}

\noindent As the statistical distance between $X$ and $X_u$ is negligible in $k$, we conclude that $X$ and $X_u$ 
are statistically indistinguishable. $\blacksquare$
\newline\newline
\noindent \textbf{Claim 2} The distribution of $Y$ and the uniform distribution $Y_u$ over binary strings of length $sk$ 
are statistically indistinguishable.
\newline\newline
\noindent For any element in the support of $Y_u$, partition it into a sequence of $s$ substrings of length $k$ 
as we do with $Y$.  Note that, for $Y$, any $y_i$ is equal to $L(x_i) = (a0 + \sum_{i=1}^{\lvert \vec{a_i} \rvert} x^i a_i)$, 
and so $Y$ is determined by the randomly chosen $a_0,...,a_{s=e-1}$ and $X$.  We first look at the probability of 
observing any string $Y$ conditioned upon a specific observation of $X$:

\[
Pr_{a_0,...,a_{s=e-1}}[Y = y_1,...,y_s \mid X = x_1,...,x_s]
\]

\noindent Fix $a_0$.  Note that, given any pairwise distinct $X = x_1,...,x_s$, it is possible to observe $y_1,...,y_s$: simply take $L$ 
as the degree-$s$ polynomial fitting $(0,a_0),(x_1,y_1),...,(x_s,y_s)$.  But (for $a_0$ fixed) $a_1,...,a_s$ determine the 
curve $L$.  Since there are $2^{sk}$ possible curves and $2^{sk}$ possible realizations of $Y$, there must exist a one-to-one 
correspondence between $a_1,...,a_s$ and $y_1,...,y_s$ given $x_1,...,x_s$.  Thus, 

\begin{align*}
Pr_{a_1,...,a_{s=e-1}}[Y = y_1,...,y_s \mid X = x_1,...,x_s \land a_0 = p_0] &= Pr[a_1,...,a_s \sim \{0,1\}^{sk}] \\
&= \frac{1}{2^{sk}}
\end{align*}

\noindent Now remove the restriction on $a_0$. 

\begin{align*}
Pr_{a_0, a_1,...,a_{s=e-1}}[Y = y_1,...,y_s \mid X = x_1,...,x_s] &= \sum_{p \in \{0,1\}^k} Pr[a_0 = p] Pr[a_1,...,a_s \sim \{0,1\}^{sk}] \\
&= \sum_{p \in \{0,1\}^k} \frac{1}{2^k} Pr[a_1,...,a_s \sim \{0,1\}^{sk}] \\
&= \frac{1}{2^{sk}}
\end{align*}

\noindent The above shows that, in fact, $Y$ and $X$ are independent.  Further, it gives us that the distribution of $Y$ is 
equivalent to the distribution $Y_u$ (the statistical distance is 0).  Thus we have that $Y$ and $Y_u$ are statistically indistinguishable. $\blacksquare$.
\newline\newline
\noindent \textbf{Claim 3} $X \mid \mid Y \mid \mid \hat{Y}$ is computationally indistinguishable from the uniform random distribution over binary strings of length $3sk$.
\newline\newline
Claim 1 and claim 2 give us that $X\mid\mid Y$ is statistically indistinguishable from the uniform random distribution over binary 
strings of length $2sk$ (else we reach contradiction).  Similarly, if we assume that $\tilde{m}$ is truly random, we may apply exactly the same 
argument to $X \mid \mid \hat{Y}$.  In truth, $\tilde{m}$ is computationally indistinguishable from 
random (by nature of $P_K$ chosen as a pseudo-random permutation and $K$ chosen randomly).  We thus have that, if we may distinguish 
$\hat{X} \mid \mid \hat{Y}$ from random, we may distinguish $\tilde{m}$ from random with precisely the same advantage by direct reduction.  
By our choice of $P_K$, we thus conclude that $\hat{X} \mid \mid \hat{Y}$ is computationally indistinguishable from the uniform distribution over strings of length $2sk$ (and where the advantage of an adversary is bounded by the maximum random indistinguishability advantage of an adversary against $P_K$).

Since $X\mid\mid Y$ is statistically indistinguishable from random, and since $X \mid \mid \hat{Y}$ is computationally indistinguishable from 
random, we are able to conclude that $X \mid \mid Y \mid \mid \hat{Y}$ is computationally indistinguishable from the uniform random distribution over binary strings of length $3sk$. $\blacksquare$.
\newline\newline
\noindent Claim 3 gives us that property 3 is satisfied.  As we have already shown that property 1 is satisfied, we conclude that the given 
distributed stegosystem is OA-DS-CHA-secret. $\square$

\subsubsection{Secrecy for Variable-Length Messages Hidden in a Distributed Manner}

Note that, for a single-block message $m$, our analysis shows that we may simply directly hide $m$ (setting $a_0 = m$) and achieve 
the same security.  In the case of our system, we instead hide a random key $K$ and the encrpted message $m$: the utility of doing so 
presents itself when we make practical considerations.  In particular, by means of the indirection $K$ provides, we may achieve a 
stronger notion of  secrecy for variable-length messages: if the PRP $P$ is applied in a cipher mode offering both forward and backward diffusion, 
a warden requires access to \textit{all} message blocks in $e$ channels in order to recover anything about the original $m$.  In contrast,
if we blindly hide just $m$, the adversary requires access to only one hidden message block in $e$ channels to leak information about $m$.

\subsubsection{An Alternative Distributed Stegosystem using Channel Identification}

In this section, we provide a modified version of the previously discussed distributed stegosystem.
This stegosystem reduces the amount of information which must be published in channels at the cost 
of an increase in the amount of public knowledge needed to encode and decode messages.
\newline\newline
The public parameters of such a system are as follows:
\begin{enumerate}
\item{A field $\mathbb{F}_q$ for $q \geq 2^k$ (where $k$ is the security parameter), say $GF(2^k)$.}
\item{A unique element of $\mathbb{F}_q$ $I_{\mathcal{C}_i}$ for each channel $\mathcal{C}_i \in [\mathcal{C}]_w$.  We call $I_{\mathcal{C}_i}$ the identifier of $\mathcal{C}_i$.}
\end{enumerate}

Using precisely the same internal functions, external functions, and embedding procedure as before, 
we give $(DSD, DSE)$ as follows:

\begin{algorithm}[H]
\caption{Distributed Steganographic Encoding Procedure}\label{1a}
\begin{algorithmic}[1]
\Procedure{$DSE$}{$r, [\mathcal{C}]_t, [h]_t, m \in \{0,1\}^k, 1 \leq e \leq t$}
\State Set $K = SecretGen(r)$.
\State Set $\vec{a_i}, \vec{b_i} = PolyGen(r,e)$.
\State Set $X = <x_i> = <I_{\mathcal{C}_i \in [\mathcal{C}]_t} \mid \forall i=1\text{ up to } t>$.
\State Set $Z = <z_i> = <I_{\mathcal{C}_i \in [\mathcal{C}]_t} \mid \forall i=1\text{ up to } t>$.
\State Set $\tilde{m} = P_K(m)$.
\For{$i$ from 1 to $t$}
\State Take $y_i \mid \mid \tilde{y_i}$ as $EvaluatePoint(K, \vec{a_i}, x_i)_y \mid \mid EvaluatePoint(\tilde{m}, \vec{b_i}, z_i)_y$
\State Set $c^{(i)} = EMBED(\mathcal{C}_i, h_i, y_i \mid \mid \tilde{y_i})$.
\EndFor
\State \textbf{return} $\{(\mathcal{C}_i, <c^{(i)}>) \mid \forall =1...t \}$
\EndProcedure
\end{algorithmic}
\end{algorithm}

\begin{algorithm}[H]
\caption{Distributed Steganographic Decoding Procedure}\label{1a}
\begin{algorithmic}[1]
\Procedure{$DSD$}{$[\mathcal{C}]_s, [h]_s, \{ (\mathcal{S}_1, <c^{(1)}>), ..., (\mathcal{S}_s, <c^{(s)}>) \}$}
\For{$i$ from 1 to $s$}
\State Take $y_i \mid \mid \tilde{y_i}$ as $EXTRACT(\mathcal{S}_i, h_i, c^{(i)})$.
\State Take as a point $(I_{\mathcal{S}_i}, y_i = f(x_i))$ and append it to $L_1$.
\State Take as a point $(I_{\mathcal{S}_i}, \tilde{y_i} = g(z_i))$ and append it to $L_2$.
\EndFor
\State Set $K = Interpolate(L_1,e)$.
\State Set $\tilde{m} = Interpolate(L_2, e)$.
\State \textbf{return} $P_K^{-1}(\tilde{m})$.
\EndProcedure
\end{algorithmic}
\end{algorithm}

\noindent This scheme essentially replaces each channel's randomly-chosen x-coordinate 
with its identifier.  Consider this scheme in the context of the OA-DS-CHA secrecy game.
In claim 2 of our proof of secrecy of the previous scheme, we established that

\[ Pr_{a_0, a_1,...,a_{s=e-1}}[Y = y_1,...,y_s \mid X = x_1,...,x_s] = 2^{-sk} \]

\noindent for pairwise distinct $x_1,...,x_s$, meaning that the collection of $y_i$ are 
distributed according to the uniform random distribution.  Likewise, in claim 3 of our proof of 
secrecy, we showed that (again, for pairwise distinct $x_1,...,x_s$), the collection of 
$\hat{y_i}$ is computationally indistinguishable from random (as a result of the random appearance 
of $P_K$).  Given a proper embed-extract procedure, the OA-DS-CHA secrecy of this modified scheme 
thus follows directly given that we have shown that inputs are indistinguishable from random.
\newline\newline
\noindent \textit{Discussion } This modified scheme certainly offers direct advantages: we only need to publish $2k$ bits 
per channel (as opposed to $3k$), and we don't have to worry at all about selecting x-coordinates.
On the other hand, this scheme raises some new concerns.  For example, 

\begin{itemize}
\item{The requirement that identifiers places a hard limit on the number of channels in the 
environment.}
\item{Required knowledge of channel identifiers \textit{may} limit the number of channels with 
which a participant may interact.}
\end{itemize}

In truth, the first is not a practical concern.  If we set $k$ large enough, we would have 
enough unique identifiers available to name all of the atoms in the visible universe.  The second, 
however, is not so directly addressed.  If we choose channel identifiers naively, a participant may need to  
store $k$ bits per channel with which he or she will interact.  We may combat this issue in practice by 
choosing identification schemes on a per-environment basis which allow identification without storage.  

For the sake of illustration, say that we wish to deploy distributed steganography in an environment 
composed entirely of TCP communication between hosts.  A simple channel identification scheme that 
does not require per-channel storage would be as follows: denote the duplex TCP channel between 
host A having 32-bit IP-address $R$ and host B having 32-bit IP-address $S$ by the identifier 
$R \mid \mid S$.  

Note that the above is a solution only for a very specific, contrived sort of environment.  It does not 
directly generalize to arbitrary environments.  Creating a truly universal approach is especially challenging, 
as a single identifier collision compromises the security of the given scheme.  We give some potential ideas 
for how to achieve `more universal' channel identification:

\begin{itemize}
\item{For physically visible channels, we may be able to use things like physical appearance to 
achieve channel verification.  For example, we might be able to achieve unique identification using, say, 
pictures and locality sensitive hashing.}
\item{If we have oracle access to all of our channels for arbitrary histories, and if underlying channel distributions have non-negligible statistical distance between them, we may use a form of distribution estimation to derive unique 
identifiers.}
\end{itemize}

\subsection{Relationship between KR-DS-CHA Secrecy and OA-DS-CHA Secrecy}

In this section, we explore the relationship between KR-DS-CHA secrecy and OA-DS-CHA secrecy.  
In particular, we show that the existence of an OA-DS-CHA-secret distributed stegosystem with specific 
properties implies the existence of a KR-DS-CHA stegosystem.  We also show that the existence of a KR-DS-CHA 
stegosystem which uses random keys implies the existence of an OA-DS-CHA stegosystem.   

\subsubsection{Constructing a KR-DS-CHA System from an OA-DS-CHA System}

We show that the existence of an OA-DS-CHA-secure distributed stegosystem following the 
operate-embed-extract paradigm with $\mathcal{D}$ fixed to the uniform distribution implies the existence 
of a KR-DS-CHA-secret distributed stegosystem.

Assume that we have such a OA-DS-CHA-secret system $S=(DSE,DSD)$.  Because $S$ is an operate-embed-extract 
stegosystem, it will have the structure 

\begin{algorithm}[H]
\caption{Distributed Steganographic Encoding Procedure}\label{1a}
\begin{algorithmic}[1]
\Procedure{$DSE$}{$r, [\mathcal{C}]_t, [h]_t, m \in \{0,1\}^k, 1 \leq e \leq t$}
\State (Interleaved operations from $F_{int}$.)
\For{$i$ from 1 to $t$}
\State $q_i = F_{ext}(...)$.
\State Set $c^{(i)} = EMBED(\mathcal{C}_i, h_i, q_i)$.
\EndFor
\State \textbf{return} $\{(\mathcal{C}_i, <c^{(i)}>) \mid \forall =1...t \}$
\EndProcedure
\end{algorithmic}
\end{algorithm}

\begin{algorithm}[H]
\caption{Distributed Steganographic Decoding Procedure}\label{1a}
\begin{algorithmic}[1]
\Procedure{$DSD$}{$[\mathcal{C}]_s, [h]_s, \{ (\mathcal{S}_1, <c^{(1)}>), ..., (\mathcal{S}_s, <c^{(s)}>) \}$}
\For{$i$ from 1 to $s$}
\State Take $q_i$ as $EXTRACT(\mathcal{S}_i, h_i, c^{(i)})$.
\EndFor
\State Set $m = $ interleaved operations from $F_{int}$, operating on all $q_1,...,q_s$. 
\State \textbf{return} $m$.
\EndProcedure
\end{algorithmic}
\end{algorithm}

We obtain a KR-DS-CHA-secret system $S'$ as follows: augment $S$ to add a key generation procedure 
which takes as a parameter $1^k$ and which assigns to each channel $\mathcal{C}_i \in [C]_w$ a key $K_{\mathcal{C}_i}$ drawn 
uniformly at random from $\{0,1\}^k$.  Next, use any pseudo-random permutation $P$ (selected also according to 
security parameter $k$) and modify $DSD$ and $DSE$:

\begin{algorithm}[H]
\caption{Distributed Steganographic Encoding Procedure}\label{1a}
\begin{algorithmic}[1]
\Procedure{$DSE$}{$r, [\mathcal{C}]_t, [h]_t, m \in \{0,1\}^k, 1 \leq e \leq t$}
\State (Interleaved operations from $F_{int}$.)
\For{$i$ from 1 to $t$}
\State $q_i = F_{ext}(...)$.
\State $q_i' = P_{K_{\mathcal{C}_i}}(q_i)$.
\State Set $c^{(i)} = EMBED(\mathcal{C}_i, h_i, q_i')$.
\EndFor
\State \textbf{return} $\{(\mathcal{C}_i, <c^{(i)}>) \mid \forall =1...t \}$
\EndProcedure
\end{algorithmic}
\end{algorithm}

\begin{algorithm}[H]
\caption{Distributed Steganographic Decoding Procedure}\label{1a}
\begin{algorithmic}[1]
\Procedure{$DSD$}{$[\mathcal{C}]_s, [h]_s, \{ (\mathcal{S}_1, <c^{(1)}>), ..., (\mathcal{S}_s, <c^{(s)}>) \}$}
\For{$i$ from 1 to $s$}
\State Take $q_i'$ as $EXTRACT(\mathcal{S}_i, h_i, c^{(i)})$.
\State Take $q_i = P_{K_{\mathcal{S}_i}}^{-1}(q_i')$.
\EndFor
\State Set $m = $ interleaved operations from $F_{int}$, operating on all $q_1,...,q_s$. 
\State \textbf{return} $m$.
\EndProcedure
\end{algorithmic}
\end{algorithm}

\noindent \textit{Security of S' } 
Any adversary against $S'$ in the KR-DS-CHA-security game must distinguish between the benign 
covertext distribution and the output of $DSE$ in the chosen channels.  By the PRF security of 
$P$, before $W$ receives the $e-1$ chosen keys, all $q_i'$ are computationally indistinguishable 
from random.  Therefore, for any channel for which $W$ does not know the key, the output of $DSE$ 
is indistinguishable from the covertext distribution by choice of $EMBED$.

Consider now the point when $W$ receives access to the $e-1$ chosen keys.  In the case that $Q$ corresponds to the 
output of $DSE$, $W$ then obtains knowledge 
of a collection of $q_i$ for $e-1$ channels; by the OA-DS-CHA secrecy of $S$ (for $S$ designed using OEE, $\mathcal{D}$ fixed 
to the uniform distribution), these $q_i$ are distributed according to the uniform distribution.  In the case that $Q$ corresponds 
to the benign covertext distributions of the environment, each apparent $\hat{q}_i'$ is distributed exactly according to $\mathcal{D}$ 
(in our case, the uniform distribution); 
applying $P^{-1}$ to these $\hat{q}_i'$ yields a collection of $\hat{q}_i$ which are random.  We thus have that the apparent $q_i$, the only 
information gained by the release of keys, does not differentiate between $Q_0$ and $Q_1$ except with negligible probability.

We thus conclude the KR-DS-CHA secrecy of $S'$ as a result of the PRF secrecy of $P$ and the original OA-DS-CHA secrecy of $S$.

\subsubsection{Constructing an OA-DS-CHA System from a KR-DS-CHA System}

We now show that the existence of a KR-DS-CHA-secret distributed stegosystem with keys chosen uniformly at random 
implies the existence of an OA-DS-CHA-secure distributed stegosystem.
\newline\newline
Say we have such a KR-DS-CHA-secret distributed stegosystem $S=(DSE,DSD)$.  We may directly construct an 
OA-DS-CHA-secret system $S'=(DSE',DSD')$ as so:

\begin{algorithm}[H]
\caption{Distributed Steganographic Encoding Procedure}\label{1a}
\begin{algorithmic}[1]
\Procedure{$DSE'$}{$r, [\mathcal{C}]_t, [h]_t, m \in \{0,1\}^k, 1 \leq e \leq t$}
\State Run the key generation procedure for the channels in $[\mathcal{C}]_t$.
\State Publish each key in its respective channel using $EMBED$; update the histories $[h]_t'$ to reflect this.
\State Execute and return from $DSE_{\vec{K}}(r, [\mathcal{C}]_t, [h]_t', m \in \{0,1\}^k, 1 \leq e \leq t)$.
\EndProcedure
\end{algorithmic}
\end{algorithm}

\begin{algorithm}[H]
\caption{Distributed Steganographic Decoding Procedure}\label{1a}
\begin{algorithmic}[1]
\Procedure{$DSD'$}{$[\mathcal{C}]_s, [h]_s, \{ (\mathcal{S}_1, <c^{(1)}>), ..., (\mathcal{S}_s, <c^{(s)}>) \}$}
\State Recover the keys for the channels in $[\mathcal{C}]_s$ using $EXTRACT$ on $\{ (\mathcal{S}_1, <c^{(1)}>), ..., (\mathcal{S}_s, <c^{(s)}>) \}$.
\State Move the portion of the covertexts $c^{(1)}...c^{(s)}$ used to recover keys into the histories $[h]_s$, obtaining 
\State $[h]_s'$ and $c^{(1)'}...c^{(s)'}$.
\State Execute and return from $DSD_{\vec{K}}([\mathcal{C}]_s, [h]_s', \{ (\mathcal{S}_1, <c^{(1)'}>), ..., (\mathcal{S}_s, <c^{(s)'}>) \})$.
\EndProcedure
\end{algorithmic}
\end{algorithm}

\noindent \textit{Security of S' } The security of $S'$ follows directly from the KR-DS-CHA security of $S$.  
The visible channels in the OA-DS-CHA secrecy game correspond exactly to the keys released in the KR-DS-CHA secrecy game.


\section{Alternative Assumptions for Public-key Steganographic Key Exchange}

The authors of \cite{BiglouPubKey} define the notion of a steganographic key exchange, and they give 
only a single construction based on the standard Diffie-Hellman assumption and Yung's method of probabilistic 
bias removal.  In this section, we present additional steganographic key exchange protocols based upon different 
assumptions.  In particular, we give a steganographic key exchange protocol based upon elliptic curve Diffie Hellman, 
demonstrating that, with only minor changes, standard cryptographic key exchanges may be converted to secure 
steganographic key exchanges.  We additionally demonstrate that there indeed exist key exchange protocols which may be 
applied in a black-box fashion under the operate-embed-extract framework to obtain analogous exchanges in the cryptographic 
sense: specifically, we show that a direct black box OEE application of the RLWE key exchange of Ding et. al. \cite{RLWE} yields a
quantum-safe steganographic key exchange which is secret under the decisional ring learning with errors assumption.

\subsection{Defining Security for Steganographic Protocols}

Before presenting these key exchange protocols, I would like to address a difficulty presented in \cite{BiglouPubKey}, 
namely the formal security setting given for steganographic key exchanges.  The setting the authors give is clearly 
sufficient for proving the security of such an exchange, but it does so by side-stepping the fact that a 
key exchange protocol is, more generally, a type of two-party protocol.  By defining a key exchange as a quadruple of 
algorithms over the same channel, we miss an opportunity to obtain some definition of security for general protocols.
\newline\newline
Define now our own definition of a steganographic protocol.  A \textit{steganographic protocol} is a concrete 
communications protocol coordinating steganographic primitives in order to achieve a steganographic objective.  
A steganographic protocol is executed between n parties $P_1,...,P_n$, each communicating via (not necessarily distinct)
channels $\mathcal{C}_1,...,\mathcal{C}_n$ with an initial set of globally accessible channel histories $h_1,...,h_n$.  
As parties execute the protocol, they publish messages on some subset of channels (updating the histories of the channel while doing so) 
until protocol execution is complete and the protocol produces some set of zero or more \textit{products}.  
The execution of a steganogaphic protocol results in a \textit{transcript} $\tau$ composed of the history of each concerned 
channel before and after each discrete event during protocol execution.

With respect to the security of a steganographic protocol, we may think either in the \textit{interactive} or 
\textit{non-interactive} senses.  The warden's task in either case is to (a) attempt to distinguish between 
a protocol transcript $\tau$ or a benign channel transcript $\tau'$ composed of ordinary channel messages or (b) 
attempt to distinguish one or more \textit{products} (defined by the protocol); a protocol is secure if no warden is able 
to successfully do either except with negligible probability.  While a non-interactive warden 
attempts to perform this task given only a transcript, an interactive warden is able to modify channel messages as they 
are published.
\newline\newline
Applying this definition to steganographic key exchange, we view a steganographic key exchange as a two-party steganographic 
protocol whose single product is a shared key.  In this section (as in \cite{BiglouPubKey}), we present steganographic key exchange 
protocols secure in the non-interactive sense.  (And shelf discussion of the interactive section for the final section of 
this report.)  

\subsection{Elliptic Curve Steganography}

The first steganographic key establishment protocol we give is a rather direct one based upon the elliptic curve 
Diffie-Hellman assumption; we design it using the OEE paradigm.  Existing elliptic curve key exchange protocols 
are themselves insufficient because they assume that it is valid to convey points on a public curve in the clear; 
however, this is not necessarily true for a covert 
communication protocol because simply the fact that the bits conveyed satisfy a curve equation is enough to 
inspire suspicion.  As such, the modification we provide is simple and minor (namely, we 
simply show that it suffices to convey only the x coordinates of points) but necessary for our purposes.
\newline\newline
\noindent \textit{The protocol } Let party $A$ be the protocol initiator.  Let $B$ be the 
responder.  Choose the following as public protocol parameters:
\begin{itemize}
\item{Two public channels $\mathcal{C}_A$ and $\mathcal{C}_B$, respectively the channels to 
be used by $A$ and $B$.  We also assume public knowledge of $h_A^{(i)}$ and $h_B^{(i)}$, respecitvely 
the channel histories of channel A and channel B after the $i$th message is transmitted in the protocol.}
\item{A valid $EMBED/EXTRACT$ scheme for both $\mathcal{C}_A$ and $\mathcal{C}_B$ with the target input 
distribution $\mathcal{D}$ chosen the uniform distribution over binary strings of length $r$.}
\item{Secure curve parameters $(f,a,b,G,n,h)$ for a binary field over $GF(2^r)$, $r \geq k + 5$, where
$k$ is the security parameter.  As usual, choose the generator $G$ such that the cofactor is small, say 
$h \leq 4$.}
\end{itemize}

\noindent The role of $A$ (initiator) in the protocol is executed as follows:
\begin{enumerate}
\item{Choose $a \in [1,...,n-1]$ uniformly at random.}
\item{Obtain curve point $(x_A, y_A) = aG$.}
\item{Publish $\hat{u}=EMBED(\mathcal{C}_A, h_A^{(0)},x_A)$ in channel $\mathcal{C}_{A}$, updating 
public channel histories to $h_{A}^{(1)}$ and $h_B^{(1)}$.}
\item{Receive $\hat{v}=EMBED(\mathcal{C}_B, h_B^{(1)},x_B)$ on channel $\mathcal{C}_{B}$.}
\item{Obtain $x_B = EXTRACT(\mathcal{C}_B, h_B^{(1)}, \hat{v})$.}
\item{Solve the quadratic equation $y_B^2=x_B^3 + ax_B + c$ in the chosen field $GF(2^r)$.  
Explicitly, $y_B = (x_B^3 + ax_B + b)^{2^{r-1}}$.}
\item{Obtain curve point $(x_K,y_K) = a(x_B, y_B)$, and take $x_K$ as the shared key.}
\end{enumerate}

\noindent The role of $B$ (responder) in the protocol is executed as

\begin{enumerate}
\item{Choose $d \in [1,...,n-1]$ uniformly at random.}
\item{Obtain curve point $(x_B, y_B) = dG$.}
\item{Receive $\hat{u}=EMBED(\mathcal{C}_A, h_A^{(0)},x_A)$ on channel $\mathcal{C}_{A}$.}
\item{Publish $\hat{v}=EMBED(\mathcal{C}_B, h_B^{(1)},x_B)$ in channel $\mathcal{C}_{B}$, updating 
public channel histories to $h_{A}^{(2)}$ and $h_B^{(2)}$.}
\item{Obtain $x_A = EXTRACT(\mathcal{C}_A, h_A^{(0)}, \hat{u})$.}
\item{Solve the quadratic equation $y_A^2=x_A^3 + ax_A + c$ in the chosen field $GF(2^r)$.  
Explicitly, $y_A = (x_A^3 + ax_A + b)^{2^{r-1}}$.}
\item{Obtain the curve point $(x_K, y_K) =d(x_A,y_B)$.}
\end{enumerate}

\noindent \textbf{Proof of correctness }  We prove that, at the end of protocol execution, 
both $A$ and $B$ have knowledge of $x_K$, the protocol product.  This proof makes use of the following fact:
\newline\newline
\noindent \textbf{Fact 1 (F1): } Every quadratic equation of the form $y^2=w$ has a 
unique solution in $GF(2^r)$.  

We briefly note that this fact follows from the property that every element in 
$GF(2^r)$ is a quadratic residue.  Consider any $q \in GF(2^r)$ and  
$p=q^{2^{r-1}}$.  $p^2 = (q^{2^{r-1}})^2 = q^{2*2^{r-1}} = q^{2^r}$.  Since 
$GF(2^r)$ is cyclic, order $2^r$, $p^2 = q$.  Since the function 
$f(q) = \sqrt{q} = q^{2^{r-1}}$ is defined for all of $q$, and since it is invertible (
$f^-1(t) = t^2$), $f(q) = \sqrt{q}$ is a bijection; thus, solutions to equations of the form 
$y^2 = b$ have a unique solution given by $f(b)$.$\blacksquare$.
\newline\newline
Consider now the given steganographic key exchange protocol.  By the correctness of 
$EMBED/EXTRACT$, we know that $A$ is able to obtain $x_B$ and that $B$ is able to obtain 
$x_A$.  By F1, we have that $A$ is able to derive $y_B = (x_B^3 + ax_B + b)^{2^{r-1}}$, thus 
obtaining $B$'s chosen point $(x_B, y_B)$.  By F1, we also have that $B$ is able to derive
$y_A = (x_A^3 + ax_A + b)^{2^{r-1}}$, obtaining $A$'s choen point $(x_A, y_A)$.
	In the final step, $A$ obtains $a(x_B, y_B) = adG$, and $B$ obtains $d(x_A, y_A) = daG$. 
By the commutivity of point multiplication in elliptic curve groups, $(x_K, y_K) = adG = daG$, and so 
$B$ and $A$ agree on $x_K$.
\newline\newline
\noindent \textbf{Proof of security } We show that messages exchanged in the protocol transcript 
$\tau$ are indistinguishable from the underlying channel distributions.  Because we are operating in the 
OEE framework, and because messages exchanged hide only the $x$ coordinates $x_A$ and $x_B$, it 
is sufficient to show that $x_A$ and $x_B$ are indistinguishable from the uniform distribution over
binary strings of length $r$.  Moreover, because $x_A$ and $x_B$ are sampled in precisely the same manner,
it suffices to show that this manner of sampling produces x-coordinates indistinguishable from 
the uniform distribution.

Let $X$ be the uniform distribution over length-$r$ binary strings: for $x \in \{0,1\}^r$, 
$Pr_X(x) = \frac{1}{2^r}$.  Let $Y$ be the distribution over the x-coordinates of points 
sampled according to the manner used in the given protocol.  Note that since the curve 
is defined over $GF(2^r)$, $(x,\cdot)$ is on the curve if and only if $x \in GF(2^r)$ which 
can occur if and only if $x \in \{0,1\}^r$, so the support of the two distributions is the same.

By F1, for every $x \in GF(2^m)$, there is at most one point $(x,y)$ which lies on the curve. 
As a result, every point on the curve has a unique x-coordinate, and so the subgroup defined 
by the generator $G$ defines a subset of $\{0,1\}^m$ which may be sampled; furthermore, 
by selecting a random unique point in the subgroup with $nG$, $n$ chosen uniformaly at random 
from $1...n-1$, we are equivalently selecting uniformly at random among this subset $S \subseteq \{0,1\}^m$.

The size of $S$ is equivalent to the size of the subgroup defined by $G$:

\[ \lvert S \rvert = n - 1 = \frac{N}{h} - 1 \]

\noindent (where $N$ is the number of points on the curve and $h$ is the cofactor).  
Since we have chosen the curve such that the cofactor is less than 4, we have 

\[ \lvert S \rvert \geq \frac{N}{4} - 1 \]

\noindent Using a straight-forward application of Hasse's theorem for elliptic curves, 
we can obtain a lower bound on $N$ which we can then use to lower bound $\lvert S \rvert$:

\begin{align*}
&\lvert N - 2^r - 1 \rvert \leq 2 \sqrt{2^r} & \text{(Hasse's theorem)} \\
\Rightarrow& 2^r + 1 - N \leq 2^{r/2 + 1} \\
& -N \leq 2^{r/2} - 2^r - 1 \\
\Rightarrow& N \geq 2^r-2^{r/2}-1 \\
& N \geq 2^{r/2}(2^{r/2}-1) - 1 \\
& N \geq 2^{r/2}2^{r/2-1} - 1 \\
& N \geq 2^{r-1} - 1\\
& N \geq 2^{r-2} 
\end{align*}

\noindent We now use this to obtain a lower bound on $\lvert S \rvert$:

\begin{align*}
\lvert S \rvert &\geq \frac{N}{4} - 1 \\
&\geq \frac{2^{r-2} }{4} - 1 \\
&\geq 2^{r-5}
\end{align*}

\noindent To establish indistinguishability, we now show that the statistical distance between 
the distributions $X$ and $Y$ is small.  For any element $x \in \{0,1\}^r$, we know 
$Pr_X(x) = \frac{1}{2^r}$.  But, for $Y$, we know via the lower bound on the size of $S$ that

\begin{align*}
Pr_Y(x) = \begin{cases} 
      0 & x \not\in S \\
      \frac{1}{2^r} \leq \cdot \leq \frac{1}{2^{r-5}} & x \in S
   \end{cases}
\end{align*}

\noindent and so the statistical distance between $X$ and $Y$, $\Delta(X,Y)$ is 

\begin{align*}
\Delta(X,Y) &= sup_{x \in \{0,1\}^m}(\lvert Pr_X(x) - Pr_Y(y) \rvert) \\
&\leq \frac{1}{2^{r-5}} - \frac{1}{2 ^r} \\
&= negligible(r) \Rightarrow negligible(k)
\end{align*}

\noindent Since the statistical distance is negligible, we assume that the x-coordinate 
sampled is indistinguishable from random.  We thus conclude by the guarantees of EMBED/EXTRACT
that the transcript is indistinguishable from the underlying channel distribution.
\newline\newline
\noindent The only remaining element of security to prove is that an adversary does not 
gain knowledge of the shared key.  This follows directly from the elliptic curve Diffie-Hellman 
assumption.

\subsection{Quantum-safe Steganography: Black-box OEE Application of the RLWE Key Exchange}

In \cite{RLWE}, Ding et. al. give a relatively simple cryptographic key exchange secure 
under the ring learning with errors (RLWE) assumption.  We show that we may apply this 
exchange in a black-box fashion to obtain a steganographic key exchange which is secret 
under the decisional variant of the RLWE assumption.  One point of significance of this 
attempt is that it shows the existence of quantum-safe (under our current knowledge, of course)
steganography.
\newline\newline
\noindent \textit{The Protocol } We illustrate this protocol and its OEE application by 
explicitly enumerating its steps.  We reiterate that this protocol is identical to the one 
given in \cite{RLWE}, save for the fact that we utilize OEE during communication between initiator 
and responder.
\newline\newline
Let $A$ be the protocol initiator, communicating on channel $\mathcal{C}_A$.  
Let $B$ be the protocol responder, communicating on channel $\mathcal{C}_B$.  Pick a 
public EMBED/EXTRACT procedure $EMBED(\cdot,\cdot, c\dot)/EXTRACT(\cdot, \cdot, \cdot)$ 
for $\mathcal{C}_A$ and $\mathcal{C}_B$ which is satisfactory for $\mathcal{D}$ chosen as the 
uniform random distribution.

Choose a prime $q$, a degree $n$, a polynomial $a$, a ring $R_q=Z_q/\Phi(x)$ suitable for security 
parameter $\lambda$ (wlog, $q > 2^\lambda$) (we omit details such as the choice 
of specific prime) as a set of public parameters.  Additionally fix a sampling method.
\newline\newline
\noindent The protocol initiator performs the following steps:
\begin{enumerate}
\item{Sample two small polynomials $s_A$ and $e_A$.}
\item{Compute $p_A = as_A + 2e_I$ (let $p_A$ be represented as a list of coefficients).}
\item{Publish $\hat{u} = EMBED(\mathcal{C}_A, h_A^{(0)}, p_A)$ on channel $\mathcal{C}_A$, updating 
channel histories as usual.}
\item{Receive $\hat{v} = EMBED(\mathcal{C}_B, h_B^{(1)}, p_B\mid \mid w)$ on channel $\mathcal{C}_B$.}
\item{Obtain $p_B\mid\mid w = EXTRACT(\mathcal{C}_B, h_B^{(1)}, \hat{v})$.}
\item{Sample a small polynomial $e'_A$.}
\item{Compute $k_A = p_B s_A + 2e'_A = as_As_B + 2e_Bs_A + 2e'_A$.}
\item{Obtain key bits by applying the coefficient-wise opperation $(k_A^{(i)} + w_i \frac{q-1}{2}) \mod q \mod 2$
(eliminating error terms).}
\end{enumerate}

\noindent The protocol responder performs the following:
\begin{enumerate}
\item{Sample two small polynomials $s_B$ and $e_B$.}
\item{Compute $p_B = as_B + 2e_B$.}
\item{Receive $\hat{u} = EMBED(\mathcal{C}_A, h_A^{(0)}, p_A)$ on channel $\mathcal{C}_A$.}
\item{Obtain $p_A = EXTRACT(\mathcal{C}_A, h_A^{(0)}, \hat{u})$.}
\item{Sample a small polynomial $e'_B$.}
\item{Compute $k_B = p_A s_B + 2e'_B = as_As_B + 2e_As_B + 2e'_B$.}
\item{Obtain reconcilliation information $w$ coefficient-wise as
\begin{align*} w_i = \begin{cases} 
0 & k_B^{(i)} \in [-\frac{q}{4}, \frac{q}{4}] \\ 
1 & \text{ else}\end{cases} 
\end{align*}
}
\item{Publish $\hat{v} = EMBED(\mathcal{C}_B, h_B^{(1)}, p_B\mid \mid w)$ on channel $\mathcal{C}_B$, 
updating channel histories as usual.}
\item{Obtain key bits in the same manner as $A$, using $k_B$ instead of $k_A$.}
\end{enumerate}

\noindent The correctness of this protocol follows directly from the correctness of 
the EMBED/EXTRACT and the correctness of the original RLWE key exchange protocol.
\newline\newline
\noindent \textit{Proof of security } Note that the confidentiality of the product (the produced key) 
follows from the security of the original RLWE key exchange.  All that remains to be shown is that 
the messages exchanged during protocol execution are indistinguishable from the underlying channel distribution.
\newline\newline
\noindent \textbf{Lemma 1 } $p_A$, $p_B$, and $k_B$ are indistinguishable from polynomials chosen 
randomly from $F_q$ under the decisional RLWE assumption.
\newline\newline
That this holds for $p_A = as_A + 2e_I$ and $p_B = as_B + 2e_B$ is a direct consequence of the 
decisional RLWE assumption; the case for $k_B = as_As_B + 2e_As_B + 2e'_B$ is nearly as direct.
Since $p_A$ and $p_B$ are indistinguishable, that an adversary has knowledge of them is of no consequence.
Assume that $k_B$ is distinguishable; then there exists a polytime distinguisher $D$ which succeeds with 
non-negligible probability.  We can then construct a general-case distinguisher for the RLWE 
problem as follows:

\begin{enumerate}
\item{Receive pair $(a(x), b(x))$, where $b(x)$ is either $b(x) = as_A + 2e_A$ or a random polynomial.}
\item{Sample small polynomials $s_B, e'_B$ and compute $b'(x) = b(x)s_B + 2e'_B$.}
\item{Pass $b'(x)$ to $D$ and return the result.}
\end{enumerate}

\noindent Consider the case when $b(x)$ is random.  Then $b'(x)$ is random, and we expect 
$D$ to indicate such with non-negligible probability.  Consider the case when $b(x) = b(x) = as_A + 2e_A$.
Then $b'(x) = as_As_B + 2e_Ae_B + 2e'_B$, and we expect $B$ to distinguish as such with high probability.
Thus our general-case distinguisher also distinguishes properly with non-negligible probability, which is 
impossible under the decisional RLWE assumption.
\newline\newline
\noindent \textbf{Lemma 2 } Let $r$ be the least integer such that $2^r \geq q$.  
The coefficients of $p_A$ and $p_B$ are computationally indistinguishable 
from the uniform distribution over $\{0,1\}^{r n}$.
\newline\newline
By Lemma 1, $p_A$ and $p_B$ are indistinguishable from polynomials chosen at random from $F_q$.  
Consider the case for any polynomial sampled in the manner of $p_A$ or $p_B$.  Any 
individual coefficient must necessarily be indistinguishable from the distribution

\begin{align*}
p_{X_i}(x) = \begin{cases}
\frac{1}{q} & x \leq q \\
0 & q < x \leq 2^r
\end{cases}
\end{align*}

\noindent and must also appear to be independent.  For a truly random coefficient, the distribution 
should be 

\begin{align*}
p_{Y_i}(x) = \frac{1}{2^r}
\end{align*}

\noindent also independent.  We therefore have that, the PDF of all $n$ sampled coefficients 
takes one of two values, $P_{X}(x) \in \{0, q^{-n}\}$.  The PDF of truly random coefficients 
would be $P_{Y}(x) = \frac{1}{2^{nr}}$.  Since both $\frac{1}{2^rn} - 0$ and 
$\frac{1}{q} - \frac{1}{2^rn} = \frac{1}{2^{r(n-1)}}$ are both negligible, we conclude that 
the two distributions are statistically indistinguishable.  Further, since Lemma 1 gives us 
that the coefficients of $p_A$ and $p_B$ must be computationally indistinguishable from 
$X$, $p_A$ and $p_B$ must therefore be computationally indistinguishable from $Y$.
\newline\newline
\noindent \textbf{Lemma 3 } $w$ is indistinguishable from the uniform distribution over 
bit strings of length $n$.
\newline\newline
For a polynomial chosen at random from $F_q$, the probability that coefficient $i$ 
is in the range $[-\frac{q}{4}, \frac{q}{4}]$ (the probability that $w_i = 0$) is exactly equal to 
$\frac{\frac{q}{2}}{q-1} = \frac{q}{q-1} \frac{1}{2}$, a negligible factor off of 
$\frac{1}{2}$.  
	By Lemma 1, $k_B$ is indistinguishable from a polynomial chosen at 
random from $F_q$, and so it must hold that $w_i = 0$ with probability negligibly far from 
$\frac{1}{2}$ (and thus that $w_i=1$ with probability negligibly far from $\frac{1}{2}$).  Since 
this holds for all coordinates, the entire bit string of $w$ must be computationally indistinguishable 
from a uniformly random string of $n$ bits.
\newline\newline
Note now that the only inputs to $EMBED/EXTRACT$ are $p_A$, $p_B$, and $w$.  Since we have 
shown that these are each indistinguishable from the uniform random distribution under 
decisional RLWE, the indistinguishability of produced covertexts follows directly
from the choice of EMBED and EXTRACT.

\section{Directions for Future Work}

\subsection{Using OEE to Decouple Channels from Objectives}

Thus far, we have shown that operate-embed-extract may be applied to easily 
achieve new steganographic objectives, extend existing ones, and to easily connect techniques in standard 
cryptography to steganography.  As has been a theme, we have been able to do this without 
relying on specific channel characterizations while simultaneously not limiting things like efficiency 
and communication rate.

In the same manner in which we may use OEE to give provably secret steganographic constructions for any channel, 
we may also use the paradigm to engineer ways in which to maximize rate of secrecy and address implementation 
concerns in specific channels for any objective.  In this sub-section, we 
suggest and discuss potential future work in this specific area.

\subsubsection{Cryptography as a Channel}

In the report preceeding this one, we explored the prospect of using cryptographic primitives and protocols as steganographic channels.  In particular, we used the fact that initialization vectors (IVs) are generally expected to appear uniformly random 
in order to design and implement symmetric-key stegosystems in the tone of those 
suggested in \cite{BiglouPSS}.

We briefly show in this section that OEE allows us to generalize such an approach 
to apply for any objective achievable with an OEE scheme.  Beginning by 
converting the specific approach used in the previous report into an $EMBED/EXTRACT$
procedure applicable across objectives, we give some discussion on how to extend
this technique even further in order to apply to other primitives, taking 
garbled circuits as an example.

We note that work in this area seems to be particularly valuable because 
the ability to use cryptography for steganography translates to an ability to 
use any system implementing cryptography to obtain a system implementing steganography 
without excessive effort; this opens the door for future studies on the 
\newline\newline 
\textbf{IV Steganography: One Approach } Our previous report showed that any block cipher mode of operation  
making use of an explicitly conveyed uniformly random initialization vector 
may be used 
securely as a cover channel in the context of specific private-key 
stegosystems hiding 1 bit of secret per 1 bit of covertext.  Following our exposition of OEE, we see that something 
stronger is true: any such block cipher mode of operation may 
be used securely as a cover channel in the context of ANY steganographic 
objective which may be achieved with an OEE stegosystem while achieving 
the same rate.

Under the OEE paradigm, all that must be shown is the existence of an 
$EMBED/EXTRACT$ procedure secure for some input distribution 
$\mathcal{D}$. In the case of a channel $\mathcal{C}$ whose support consists of  
uniformly random IVs of length $b$, the identity operation satisfies this definition
when we take $\mathcal{D}$ as the uniform distribution over $b$-bit strings.  
Because the input distribution is precisely the same as the channel distribution 
in this case, indistinguishability, validity, and the 1:1 rate of this
$EMBED/EXTRACT$ procedure follow directly. 
\newline\newline
\textbf{Other Approaches }  Future work, of course, should concern 
itself with either presenting new techniques for other primitives or suggesting 
ways to extend the one suggested.  One interesting possibility which we have 
noted is the ability to apply this very technique to garbled circuits: in the 
same manner in which the identity operation serves as a satisfactory 
$EMBED/EXTRACT$ procedure for initialization vector, the same can be 
said about individual wire keys in a garbled circuit.  There are several 
possibilities for exploration here.  We briefly discuss two:

\begin{enumerate}
\item{$EMBED/EXTRACT$ procedures which hide messages across multiple wires.
One obvious method is to hide a message across some path of wires from some 
input bit to some output bit; another could be to encode messages across 
individual `levels' of the circuit.}
\item{How to balance privacy with a desire for covert communication.  Clearly,
if every key hides a message that our computational partner may obtain 
(say we are both trying to compute a function securely with our partner and 
communicate steganographically), we compromise privacy.  One area of exploration 
could involve looking at how we can balance a desire to communicate covertly at a 
high rate while maintaining some quantifiable degree of privacy with a partner.}
\end{enumerate}

\subsubsection{Natural Language Channels}

As we noted in our previous report, the value of a stegosystem employed in 
practice is intimately related to how much we can expect to be able to rely 
on the presence of specific channels we know to admit high rate.  In the 
case of cryptography as a channel, for example, a nation-state might 
simply outlaw the use of cryptography in order to prevent steganographic communication.
As such, the use of natural language as a steganographic channel seems to be an 
extremely valuable goal because it is much harder to regulate.  In this section, we discuss some difficulties and directions in this area.
\newline\newline
\noindent \textbf{Choosing Subliminal Features } As of now, the only 
truly `obvious' approach to using natural languages for steganography is the simple 
application of the proven universal embedding procedures we previously discussed.
While they guarantee security in theory, the requirement that there 
exists an oracle for the channel--essentially a natural language oracle--poses 
particular challenges.  For example, if we are to use a human as an oracle, we're 
almost guaranteed to lose with respect to the rate of secrecy possible.  For example, if 
we require our oracle to hide at the granularity of words 
(1 bit of secrecy per word), a human will quickly exhaust his or her ability to paraphrase before the 
entire message may be embedded for messages longer than a few bits.  We may, however, sample at a coarser granularity,
say at the level of sentences and paragraphs, and achieve human oracle feasibility at the cost of a lower rate.

As such, it makes sense to consider different language-based features to use 
in the context of universal approaches.  One possibility which we have considered
is the idea of hiding in the medium of language rather than the language itself.
In particular, we have considered the possibility of hiding on a per-letter basis 
in hand-written language.  If there is enough natural variance in the way a human
writes the same letter of the alphabet across multiple attempts, for example, 
we may be able to achieve both high rate (on the order of one bit 
per letter) and human oracle feasibility.  The relevance of this approach 
in modern times is dubitable, however, when we consider that people only rarely
communicate via hand-written messages.
\newline\newline
\noindent \textbf{Constructing Non-human Oracles } The previous discussion 
assumed a need to use human oracles.  It might be possible to instead develop 
non-human oracles (e.g. a paraphrasing model) which is capable of praphrasing
accurately and in high volume, but this seems especially challenging in that 
such a model would need to be virtually perfect in order to guarantee security.

In particular, an interesting task we could propose in the context of NLP which 
might address our concerns in a heuristic sense (or in an exact sense 
if we can expect perfect models) while admitting high rate is a task which 
might perhaps be described as `indexed paraphrasing': given a partial message $m$, a statement $s$, and 
an index $i$, determine the $i$th paraphrase of $s$ according to some arbitrary 
ordering of all statements with the same meaning and style as $s$ with respect to $m$.
This task seems to be extremely hard, if not impossible, and so it may be better to 
instead invest time into either constructing high-quality simple paraphrase models or 
simply determining clever language-based features to use in conjunction with 
universal embedding. 

\subsection{Steganographic Public-key Infrastructures (S-PKIs)}

Hopper and von Ahn's original paper on public-key steganography \cite{BiglouPubKey}
raises a concern regarding a topic extremely important to the practical deployment
of infrastructure making use of steganography at scale: the issue of distributing,
maintaining, and verifying steganographic keys.  In the context of steganography,
many problems relating to key management, non-repudiation, and authentication 
suddenly become hard.

As the authors note, it seems that the use of steganographic key exchange protocols 
may be necessary.  The authors note that it also appears to be necessary to 
have a `one-bit secret channel' which may be used to indicate that somebody is attempting to perform a steganographic key exchange, but this is not strictly 
true in practice.  For example, all members of the S-PKI may simply maintain 
a directory of other members and institute a policy that all members initiate 
communication using a steganographic key exchange and then use that key to indicate to either (a) abort steganographic interaction or (b) continue.

As such, one route of inquiry could explore whether it's possible to develop an enveloping procedure using steganographic key exchanges to convert a standard PKI 
into an S-PKI.  Even if we could do so, however, there are still some outlying 
questions which must be answered.  Some of these include the following:

\begin{enumerate}
\item{How do you handle malicious insiders?  An insider which leaks a public 
key may potentially reveal the entire steganographic network.}
\item{How do you induct new members into the steganographic network?}
\item{How do you remove members from the steganographic network?  Do you have to 
re-issue all keys?}
\end{enumerate}

\noindent The obvious remedy required by the above is a more nuanced approach 
to trust and the manner in which it is verified and acted upon: future work would 
do well to focus on this topic.

While steganography is both interesting and potentially of great value to the world, 
our lack of means to manage steganographic interaction at scale could possibly 
be the largest barrier to its realization.  As such, future work should 
certainly seek to address these gaps.

\section{Conclusion}

This report has sought to explore steganography beyond the scope of private-key steganography, 
introducing a paradigm through which we may consider any number of other goals.  Beginning with an exploration of the 
current formalism for public-key steganography given in \cite{BiglouPubKey}, this report attempts to 
condense existing work into a unifying design paradigm that admits provably secret constructions while 
allowing for both universal constructions and constructions with channel-specific optimizations.  We 
show the utility of this paradigm by using it to design alternative public-key constructions 
and even achieve a new goal, distributed steganography.  Following a presentation of this 
paradigm and its applications, we conclude with a general discussion of (1) how this paradigm 
may be further applied to address issues in practice and (2) other issues preventing widespread 
use of steganography.




\bibliographystyle{acm}
\bibliography{report}

\end{document}